\documentclass[a4paper,twocolumn,superscriptaddress]{revtex4-2}
\usepackage{amsmath}
\usepackage{amssymb}
\usepackage{graphicx}
\usepackage{xcolor}

\newcommand{\nn}{\nonumber\\}

\begin{document}
\title{Signatures of cooperative emission in photon coincidence: 
Superradiance versus measurement-induced cooperativity}
\author{Moritz Cygorek}
\affiliation{SUPA, Institute of Photonics and Quantum Sciences, Heriot-Watt University, Edinburgh EH14 4AS, United Kingdom}
\author{Eleanor D. Scerri}
\affiliation{SUPA, Institute of Photonics and Quantum Sciences, Heriot-Watt University, Edinburgh EH14 4AS, United Kingdom}
\author{Ted S. Santana}
\affiliation{National Physical Laboratory (NPL), Hampton Road, Teddington, TW11 0LW, United Kingdom}
\author{Zhe X. Koong}
\affiliation{SUPA, Institute of Photonics and Quantum Sciences, Heriot-Watt University, Edinburgh EH14 4AS, United Kingdom}
\author{Brian D. Gerardot}
\affiliation{SUPA, Institute of Photonics and Quantum Sciences, Heriot-Watt University, Edinburgh EH14 4AS, United Kingdom}
\author{Erik M. Gauger}
\affiliation{SUPA, Institute of Photonics and Quantum Sciences, Heriot-Watt University, Edinburgh EH14 4AS, United Kingdom}

\begin{abstract}
Indistinguishable quantum emitters confined to length scales smaller than 
the wavelength of the light become superradiant. 
Compared to uncorrelated and distinguishable emitters, superradiance results 
in qualitative modifications of optical signals such as photon coincidences. 
However, recent experiments 
revealed that similar signatures can also be
obtained in situations where emitters are too far separated to be 
superradiant if correlations between emitters are induced by the wave function
collapse during an emission-angle-selective photon detection event.
Here, we compare two sources for cooperative emission,
superradiance and measurement-induced cooperativity, 
and analyze their impact on time-dependent optical signals. 
We find that an anti-dip in photon coincidences at 
zero time delay is a signature of inter-emitter correlations in general but
does not unambiguously prove the presence of superradiance. 
This suggests that photon coincidences at zero time delay alone are not 
sufficient and time-dependent data is necessary to clearly 
demonstrate a superradiant enhancement of the spontaneous radiative decay rate.
\end{abstract}

\maketitle

\section{Introduction}

Spontaneous photon emission is one of the most elementary processes
in quantum physics\cite{Einstein,Weisskopf}. 
In many situations, it is appropriately described by 
the conversion of excitations of quantum emitters into photons 
with some fixed rate $\Gamma$ depending on the particular emitter\cite{Einstein}.
However, a closer look reveals that even spontaneous emission can reveal
interesting insights into fundamental aspects of quantum mechanics.
For example, it has been realized that radiative decay not only
depends on the emitters themselves but also on their photonic environment.
Drexhage \emph{et al.}\cite{Drexhage} famously observed substantial 
changes of photon emission rates when emitters are placed close to a mirror. 
Nowadays, photonic structures like waveguides\cite{Waks2018,Joens2017}, 
optical microcavities\cite{Thomas2021,PI_singlephoton}, 
or photonic crystals\cite{Fox_Purcell,photonic_crystal3D}  
have become key elements of
solid state quantum devices relying on efficient photon extraction 
via the Purcell effect
or on spectral filtering via resonances in the photonic 
environment\cite{Iles-Smith2017,delVallePRL2012,delValle2011,Tim_BiexcitonCascade_PRB,Tim_DifferentTypes}.

Moreover, even the emission into free space can reveal intricate effects of
cooperative emission when multiple indistinguishable quantum emitters 
are involved, such as in the case of superradiance\cite{Dicke, Haroche, Bradac2017}: 
On a semiclassical level, superradiance
can be understood by the fact that the spontaneous emission rate of a
single quantum emitter $\gamma$ is proportional to the 
modulus square of the transition dipole $\mathbf{d}$.
Confining $N$ emitters to volumes smaller 
than the wavelength of the light, so that the light field effectively
interacts with a single large dipole $\mathbf{D}=N\mathbf{d}$, hence yields a
spontaneous emission rate of up to $N^2 \gamma$ as opposed to the emission from
$N$ individual emitters, each with rate $\gamma$.
A more detailed quantum mechanical treatment\cite{Dicke, Haroche} reveals 
emission to take place via a cascade through the Dicke ladder, a set
of strongly correlated states with excitations equally distributed across 
many emitters.
The superextensive light-matter coupling is also the reason why the reverse
process, superabsorption\cite{superabsorption_Higgins2014, 
superabsorption_Yang2021}, has been proposed for applications, e.g., 
in quantum batteries\cite{quantum_battery}. 

Even though applications typically rely on cooperative effects 
in the large-$N$ limit, 
experiments can also provide valuable insights for 
samples with only $N=2$ or $N=3$ emitters as these often facilitate
a direct control, e.~g., varying the degree of indistinguishability by 
tuning emitters in and out of resonance\cite{Waks2018,Gammon2019,SciAdvCoop}.
In the low-$N$ limit, photon coincidence measurements are particularly useful 
because the violation of the upper bound of zero-delay coincidences 
$g^{(2)}(0) \le (N-1)/N$ for uncorrelated emitters%
\footnote{
This is because in absence of correlations 
$\langle \sigma^+_i \sigma^+_j\sigma^-_j\sigma^-_i\rangle 
=\langle \sigma^+_i \sigma^-_i\rangle\langle\sigma^+_j\sigma^-_j\rangle 
=n_i n_j$ for $i\neq j$ and 
$\langle \sigma^+_i \sigma^+_j\sigma^-_j\sigma^-_i\rangle=0$ for $i=j$.
Hence $G^{(2)}(0)=\sum_{i,j\neq i} n_i n_j=I^2(0)- \sum_i n^2_i$
and $g^{(2)}(0)=G^{(2)}(0)/I^2(0)=
1-\big(\sum_i n^2_i\big)/\big(\sum_i n_i\big)^2$.
This expression is maximal for equal $n_i$, for which 
$g^{(2)}(0)=1-N/N^2$.
} 
is clear proof for inter-emitter correlations (cf.~Appendix~\ref{app:involveCor}).
For example, values of $g^{(2)}(0)\approx 1$ have been used as the main 
piece of evidence for superradiance of semiconductor quantum dots coupled 
to a nanophotonic waveguide in the case of $N=2$ emitters\cite{Waks2018}.
Similarly, for $N=3$ quantum dots, values exceeding $g^{(2)}(0)>1$ 
have been reported\cite{Gammon2019}.

Recently, the radiation pattern of light scattering at two identical 
trapped ions\cite{Schmidt-KalerPRL,Schmidt-Kaler2022}, 
has added yet another dimension to the discussion of cooperative emission.
There, photon coincidences $g^{(2)}(0)$ have been demonstrated to 
exceed or fall below the expected value of $1/2$ 
for two identical uncorrelated emitters, depending on the detection angle. 
This can be explained by a measurement-induced preparation of a 
correlated Dicke-like state by the first photon detection event,
which shapes the radiation pattern for the subsequent emission of a
second photon\cite{Walther,Zanthier_noninteracting}.
It is noteworthy that these results 
are also found when the separation of the emitters
exceeds the wavelength of the light, where superradiance, as discussed by
Dicke\cite{Dicke}, is not expected. 

We have recently demonstrated a solid-state quantum device where two 
semiconductor quantum dots (QDs) can be electrically tuned into 
resonance\cite{SciAdvCoop}.
This device operates in a similar regime as the experiments on trapped ions 
in that the emitters are
spectrally indistinguishable but the spatial separation exceeds the value 
for which superradiance is expected. 
In contrast to Refs.~\cite{Schmidt-KalerPRL,Schmidt-Kaler2022}, 
we additionally investigated temporal aspects such as
free radiative decay for situations corresponding to distinguishable and
indistinguishable emitters, respectively, using different driving conditions
like continous pumping and pulsed excitation\cite{SciAdvCoop}. 
Detecting photons in the direction perpendicular to the plane containing
the QDs, we indeed found signatures in photon coincidences 
resembling those expected from superradiance, 
such as values of $g^{(2)}(0)> 1/2$, 
but at the same time no evidence of superradiant rate enhancement was
observed in the free radiative decay.

These observations raise important conceptual questions:
How exactly is the physical situation in 
Refs.~\cite{Schmidt-KalerPRL,Schmidt-Kaler2022} and \cite{SciAdvCoop} 
related to superradiance? If an anti-dip in $g^{(2)}(\tau)$ with
$g^{(2)}(0)$ exceeding the limit for independent emitters is not a
unique signature of superradiance, how can both situations be distinguished
by measurement?

While many aspects of superradiance, measurement-induced coherence, and 
interference of light emitted from quantum emitters at fixed positions
have been thoroughly investigated (see, e.g., Refs.~\cite{Walther,KeitelPRL,FicekBook}), 
most previous works either neglect dephasing or primarily focus on static quantities such as 
 photon coincidences $g^{(2)}(0)$ at zero delay time only.
To obtain a realistic description of cooperative emission including 
coherent and time-dependent driving as well as unavoidable dephasing  in solid-state systems, 
we here rephrase the emission dynamics
employing an open quantum systems framework for the density matrix of the emitter 
system. This allows us to treat superradiance as well as  measurement-induced 
cooperative photon emission within a single framework, and to discuss 
common features as well as differences. 

Before presenting our main results relating to time-dependent optical signals, 
we introduce our framework through a detailed pedagogical derivation, also applying 
it to elementary examples and recovering well-known limiting cases~\cite{Dicke,FicekBook,KeitelPRL}.
To preclude confusion, here, we stick to definitions where 
we use the term ``cooperative emission'' to
refer to general situations where more than one emitter is involved in a 
single photon emission process, while we reserve ``superradiance''
exclusively 
for situations where cooperative emission additionally leads to an increase 
of the overall radiative decay rate, i.e. to enhanced radiance. 
In these terms, Dicke superradiance and measurement-induced cooperativity by 
emission-angle-selective measurement can be viewed as two different instances 
of cooperative emission, even though the latter does not show any
superradiant rate enhancement.

Once these concepts have been clarified, we use our framework to present
numerical and analytical calculations of time-dependent photon coincidences 
for continuously pumped emitters as well as for emitters under pulsed driving,
while simultaneously properly accounting for effects of dephasing.
Our results elucidate how signatures of cooperative emission 
manifest in experimentally relevant situations, and highlights pitfalls relating to
the correct interpretation of measured data. As a key insight we find that, in the presence
of dephasing, the photon coincidence trace for measurement-induced cooperativity 
is qualitatively remarkably similar to that of superradiant decay. This reflects the fact that both rely on 
the presence of correlations between emitters in general rather than being an unequivocal 
indicator of a superradient decay rate enhancement.
Similarly, care has to be taken when interpreting time-integrated photon
coincidences after pulsed driving: for single and indistinguishable emitters,
time-integrated coincidences are found to have the same value as the 
corresponding zero-delay coincidences $g^{(2)}(0)$ for incoherently pumped
emitters, but the link between these two quantities generally does not 
carry over to cooperatively emitting quantum emitters.

This article is structured as follows: First we (re-)derive established
results for radiative decay and photon coincidences from two emitters 
with identical dipoles
in the cases of distinguishable and superradiant emitters.
We then generalise the treatment to obtain a framework in which general
photon emission as well as the effects of angle-resolved detection can 
be discussed, which naturally leads 
to the observation of cooperative emission due to selective measurement.
Finally, we calculate concrete optical signals such as the full delay-time
dependent photon coincidences $g^{(2)}(\tau)$ in this regime under the
assumption of incoherent continuous driving as well as 
time-integrated coincidences for a system of emitters driven by short 
laser pulses.

\section{Radiative decay of distinguishable and superradiant emitters}
Throughout this article, we consider the case of two emitters located
at positions $\mathbf{r}_1=-\mathbf{r}/2$ and 
$\mathbf{r}_2=\mathbf{r}/2$, respectively.
Modelling the $i$-th emitter
as a two-level system with ground and excited
states $|g_i\rangle$ and $|e_i\rangle$, respectively,
and introducing operators
$\sigma^+_i=|e_i\rangle\langle g_i|$ and $\sigma^-_i=|g_i\rangle\langle e_i|$,
the total Hamiltonian of the emitter system coupled to the light field modes
is
\begin{subequations}
\label{eq:hamil}
\begin{align}
H=&H_0+ H_I,
\\
H_0=&\sum_{i=1,2}\hbar\omega_i  \sigma^+_i \sigma^-_i
+\sum_{\mathbf{k},\lambda} \hbar\omega_{\mathbf{k}} 
a^\dagger_{\mathbf{k},\lambda} a_{\mathbf{k},\lambda},
\\
H_I=&\sum_{\mathbf{k},\lambda} 
\big( h_{\mathbf{k},\lambda} a^\dagger_{\mathbf{k},\lambda}
+h_{\mathbf{k},\lambda}^\dagger a_{\mathbf{k},\lambda}\big),
\\
h_{\mathbf{k},\lambda}=&\hbar 
\big( g_{1,\mathbf{k},\lambda} \, e^{i\mathbf{k}\cdot \mathbf{r}/2}\sigma^-_1
+g_{2,\mathbf{k},\lambda} \, e^{-i\mathbf{k}\cdot \mathbf{r}/2}\sigma^-_2 \big),
\end{align}
\end{subequations}
where $a^\dagger_{\mathbf{k},\lambda}$ and $a_{\mathbf{k},\lambda}$ 
are creation and annihilation operators of photons with wave vector 
$\mathbf{k}$ and polarization $\lambda$, 
$\hbar\omega_{\mathbf{k}}$ is the energy of the respective photon mode, and
$\hbar\omega_i$ is the fundamental transition energy of the $i$-th emitter.
Here, we assume identical light-matter coupling 
strengths $g_{1,\mathbf{k},\lambda}=g_{2,\mathbf{k},\lambda}=
g_{\mathbf{k},\lambda}$ 
for both emitters and all photon modes $\mathbf{k}$, where
$g_{\mathbf{k},\lambda}= 
(\hat{\mathbf{d}}\cdot\mathbf{e}_{\mathbf{k},\lambda})^2 g$ 
with constant $g$, normalized direction of the dipole $\hat{\mathbf{d}}
=\mathbf{d}/|\mathbf{d}|$, and
polarization vector $\mathbf{e}_{\mathbf{k},\lambda}$.
For a more convenient notation, we henceforth drop the polarization index 
$\lambda$ unless necessary.

If the emitters are spectrally distinguishable, i.e., 
there is vanishing overlap between the spectral lines at 
$\hbar\omega_1$ and $\hbar\omega_2$,
radiative decay can be described by non-degenerate perturbation theory 
using Fermi's golden rule, which predicts a decay rate
\begin{align}
\gamma_{\textnormal{i} \to \textnormal{f}} 
=&\frac{2\pi}{\hbar} \sum_{\mathbf{k}} \big|\langle \textnormal{f}|
h_\mathbf{k}|\textnormal{i}\rangle \big|^2
\delta(E_{\textnormal{i}}-E_{\textnormal{f}}
-\hbar\omega_{\mathbf{k}}),
\end{align}
where $|\textnormal{i}\rangle$ and $|\textnormal{f}\rangle$ are the initial 
and final states of the decay process, which are eigenstates
of the unperturbed problem $H_0$,
and $E_\textnormal{i}$ and $E_\textnormal{f}$ are the corresponding energies.
For two distinguishable emitters, the energy eigenstates are product states
of the emitters in ground or excited states
$|e_1,e_2\rangle$, $|e_1,g_2\rangle$, $|g_1,e_2\rangle$, and $|g_1,g_2\rangle$.

\begin{figure}
\includegraphics[width=\linewidth]{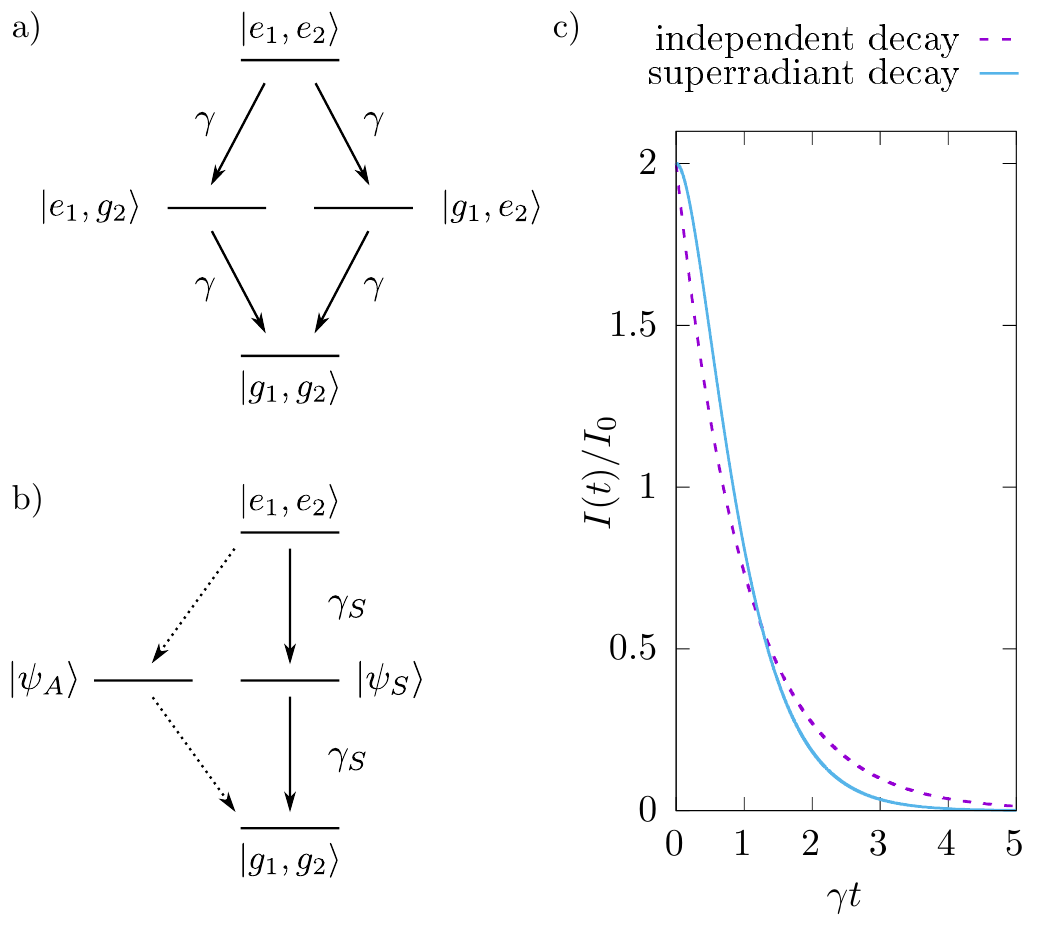}
\caption{\label{fig:ind_sup}Level scheme for radiative decay in
distinguishable (a) and superradiant (b) emitters. For ideal superradiance,
$\gamma_S=2\gamma$ while transitions involving the antisymmetric Dicke state
are completely dark.
c) Time dependence of emitted intensity for cases (a) and (b) assuming 
an initially prepared doubly excited state $|e_1,e_2\rangle$.
}
\end{figure}

Assuming a flat photon density of states
$D(E)=\sum_{\mathbf{k},\lambda}\delta(E-\hbar\omega_{\mathbf{k}})
(\hat{\mathbf{d}}\cdot \mathbf{e}_{\mathbf{k},\lambda})^2
=D$ within the range of the relevant energies,
the radiative decay rates for all processes where one excitation
is emitted as a photon, as depicted in Fig.~\ref{fig:ind_sup}a, are identical 
$\gamma=2\pi \hbar g^2 D$.

For spectrally indistinguishable emitters with $\omega_1=\omega_2$, 
non-degenerate perturbation theory no longer applies and the degeneracy 
has to be addressed explicitly.
An important special case is the superradiant regime, 
where the distance between emitters is much smaller than the wavelength of 
the light $\mathbf{k}\cdot\mathbf{r}\approx 0$. 
Then, the phase factors in the interaction
term $h_\mathbf{k}$ are $e^{\pm i\mathbf{k}\cdot\mathbf{r}/2}\approx 1$
and one can replace $h_\mathbf{k}=h_S$ with
\begin{align}
h_S=&\hbar \sqrt{2} g \sigma_S^-,
\label{eq:hs}
\end{align}
with
\begin{align}
\sigma^-_{S/A}=&\frac 1{\sqrt{2}}\big( \sigma^-_1 \pm \sigma^-_2\big) 
=
|g_1,g_2\rangle\langle \psi_{S/A}| + |\psi_{S/A}\rangle\langle e_1,e_2|,
\end{align}
where
$|\psi_{S/A}\rangle=\frac 1{\sqrt{2}}\big(
|e_1,g_2\rangle \pm |g_1,e_2\rangle\big)$
are the symmetric and antisymmetric Dicke states, respectively.
As the antisymmetric state decouples from the dynamics, 
non-degenerate perturbation theory can now be applied to the transitions
between the remaining three-level system. With Fermi's golden rule, one finds 
a cascade of transitions through the 
symmetric Dicke state $|\psi_S\rangle$, as depicted in 
Fig.~\ref{fig:ind_sup}b,
with rates $\gamma_S=2\gamma$.
This rate $\gamma_S$ is enhanced by a factor of 2 with respect
to the radiative decay rate of a single emitter $\gamma$,
originating from the enhanced dipole (by a factor of $\sqrt{2}$) in the
interaction $h_S$ in Eq.~\eqref{eq:hs}, which is a manifestation of the 
cooperation of both emitters in both emission processes. 
Note that this enhancement affects only the rate for individual transitions, 
while the overall emission rate also depends on the number of decay
channels. As there are two channels for the first photon emission
in the situation of two distinguishable emitters, 
the overall rate for the emission of a first photon is identical 
to that in the superradiant case with only one channel 
at twice the rate. It is the emission of the second photon, where in both cases
only a single channel exists, that the superradiant rate enhancement 
leads to overall increased photon emission.

The cascaded emission through the superradiant three-level system also leads
to a distinct non-exponential dynamics of the emitted intensity after
excitation of the doubly excited state as depiced in Fig.~\ref{fig:ind_sup}c
(cf. Appendix~\ref{app:pulsed_sel} or Ref.~\onlinecite{Haroche} for
explicit expressions).

Finally, to assess signatures of superradiance on optical signals, 
the photon detection process has to be modelled. 
A point-like detector in the far field at a displacement 
$\boldsymbol{\mathcal{D}}$ with respect to the center of the emitters 
picks up only photons with a fixed wave vector $\mathbf{k}$ 
whose direction is parallel to $\boldsymbol{\mathcal{D}}$ and whose magnitude 
is determined by the detected energy $\hbar\omega=\hbar c|\mathbf{k}|$. 
The detected intensity signal is given by
\begin{align}
I_\mathbf{k}(t)=&\frac{1}{\Delta \tau_M}
\langle a^\dagger_\mathbf{k}(t) a_\mathbf{k}(t)\rangle,
\end{align}
where $\Delta \tau_M$ is a characteristic timescale of the measurement,
which depends on the detector (cf. discussion in Appendix~\ref{app:krate}).
Then, the time integral $\int_{t_0}^t dt'\,I_\mathbf{k}(t')$ yields the
expectation value of the number of clicks on the detector from time $t_0$
to time $t$. 

A finite-size detector is described by a collection of point-like 
detectors using the mask function $\eta_{\mathbf{k}}$, which is 
$\eta_{\mathbf{k}}=1$ for wave numbers $\mathbf{k}$ that are picked up
by the detector and $\eta_{\mathbf{k}}=0$ otherwise.
The corresponding intensity signal is 
\begin{align}
I(t)=&\sum_{\mathbf{k}}\frac{\eta_{\mathbf{k}}}{\Delta \tau_M}
\langle a^\dagger_{\mathbf{k}}(t)a_\mathbf{k}(t)\rangle
\end{align}
Similarly, photon coincidences are given by
\begin{align}
\label{eq:G2general}
G^{(2)}(t,\tau)=&\sum_{\mathbf{k}\mathbf{k}'}
\frac{\eta_\mathbf{k}\eta_{\mathbf{k}'}}{\Delta \tau_M^2}
\langle a^\dagger_{\mathbf{k}}(t) a^\dagger_{\mathbf{k'}}(t+\tau)
a_{\mathbf{k}'}(t+\tau)a_{\mathbf{k}}(t)\rangle,
\\
g^{(2)}(t,\tau)=&\frac{G^{(2)}(t,\tau)}{I(t)I(t+\tau)},
\end{align}
for unnormalized and normalized photon coincidences, respectively.

In Appendix \ref{app:krate}, we derive in detail
how the photon emission can be
expressed in terms of the state of the emitter system for cases of
distinguishable and indistinguishable emitters.
Defining the occupations of the states $|e_1,e_2\rangle$, $|e_1,g_2\rangle$,
$|g_1,e_2\rangle$, and $|\psi_S\rangle$, 
as $n_{e_1,e_2}$, $n_{e_1,g_2}$, $n_{g_1,e_2}$, and $n_S$,
respectively, the intensities from distinguishable and superradiant emitters
are
\begin{align}
I_{\textnormal{dist}}=& I_0
\sum_{i=1,2}\langle \sigma^+_i\sigma^-_i\rangle
=I_0( 2n_{e_1,e_2} + n_{e_1,g_2} + n_{g_1,e_2} ),
\label{eq:Idist}
\\
I_{\textnormal{sup}}= & 2 I_0  
\langle \sigma^+_S\sigma^-_S\rangle=2I_0( n_{e_1,e_2} + n_S ),
\label{eq:Isup}
\end{align}
respectively, where $I_0=\sum_{\mathbf{k}}\eta_{\mathbf{k}} 
\gamma_{\mathbf{k}}^{\textnormal{single}}$ and
$\gamma_{\mathbf{k}}^{\textnormal{single}}={2\pi}{\hbar}g^2_\mathbf{k} 
\delta(\hbar\omega_\mathbf{k} - \hbar\omega)$ is the rate of photon emission 
from a single emitter into the photon mode with wave vector $\mathbf{k}$ 
derived in Appendix~\ref{app:krate}. 

The corresponding coincidences are
\begin{align}
\label{eq:G2dist}
G^{(2)}_{\textnormal{dist}}(t,\tau)=& I_0^2 
\sum_{i,j=1,2}\langle \sigma^+_i(t) \sigma^+_j(t+\tau)
\sigma^-_j(t+\tau)\sigma^-_i(t)\rangle,
\\
G^{(2)}_{\textnormal{sup}}(t,\tau)=& 4I_0^2\,
\langle \sigma^+_S(t) \sigma^+_S(t+\tau) 
\sigma^-_S(t+\tau)\sigma^-_S(t)\rangle. 
\label{eq:G2sup}
\end{align}

The normalized zero-delay coincidences $g^{(2)}(t,0)$ 
can be obtained noting that for
the distinguishable case $\sigma^-_i \sigma^-_i=0$ while 
$\sigma^-_{j\neq i} \sigma^-_i = |g_1,g_2\rangle\langle e_1,e_2|$, so that
\begin{align}
g^{(2)}_{\textnormal{dist}}(t,0)=&
\frac{ 2n_{e_1,e_2}(t) }{\big( 2n_{e_1,e_2}(t) + n_{e_1,g_2}(t) 
+ n_{g_1,e_2}(t) \big)^2},
\end{align}
which, for initially uncorrelated emitters with excited state populations 
$n_1$ and $n_2$, respectively, becomes
\begin{align}
g^{(2)}_{\textnormal{dist}}(t,0)
=&\frac{2n_1(t) n_2(t)}{ \big( n_1(t) + n_2(t) \big)^2 } \le \frac 12,
\label{eq:glimit}
\end{align}
where $g^{(2)}_{\textnormal{dist}}(t,0)=1/2$ for equally excited
emitters $n_1(t)=n_2(t)$ with identical dipoles.

Eq.~\eqref{eq:glimit} sets the limit for photon coincidences from two
independent emitters without involvement of correlations between the emitters,
irrespective of the driving or other system parameters.
As discussed in more detail in Appendix~\ref{app:involveCor},
a violation of Eq.~\eqref{eq:glimit} 
constitutes a clear signature
of cooperative effects, which requires emitters to be correlated at some
point during the photon emission.

For superradiant emitters, 
$\sigma^-_S \sigma^-_S=|g_1,g_2\rangle\langle e_1,e_2|$, 
so the zero-delay coincidences are
\begin{align}
\label{eq:g2sup0}
g^{(2)}_{\textnormal{sup}}(t,0)=&
\frac{ n_{e_1,e_2}(t) } {\big(n_{e_1,e_2}(t)+n_S(t)\big)^2},
\end{align}
which, for initially uncorrelated and equally occupied emitter states
becomes 
$g^{(2)}_{\textnormal{sup}}(t,0)=1$, 
a value that is twice as large as the limit Eq.~\eqref{eq:glimit}
for emission without cooperative effects.
\begin{figure*}
\includegraphics[width=0.95\linewidth]{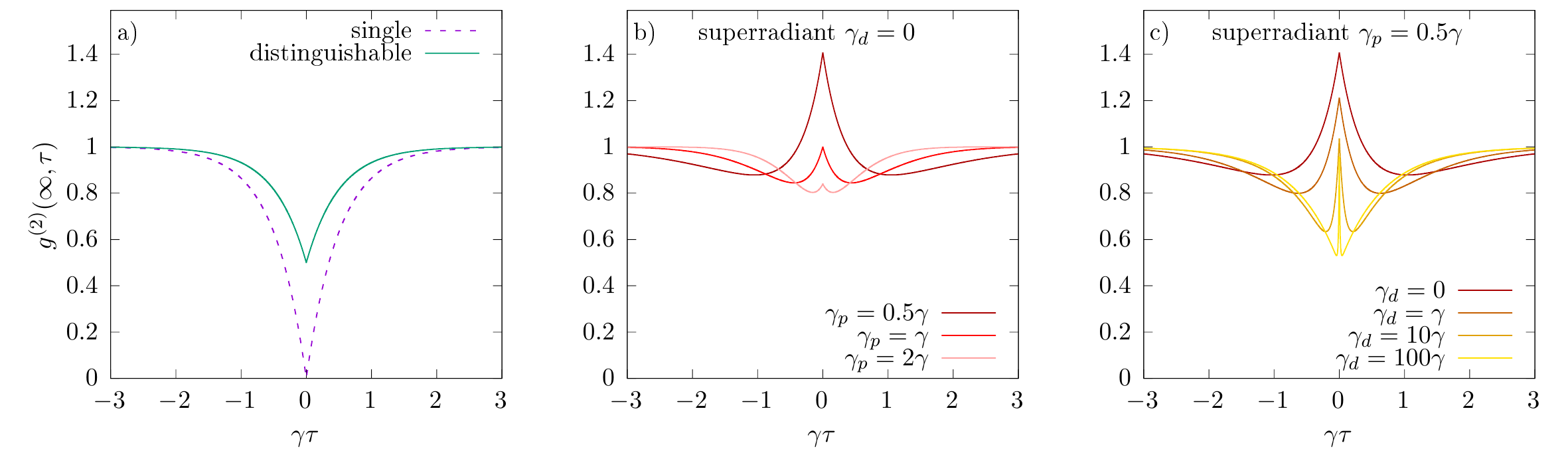}
\caption{\label{fig:dip}Photon coincidences $g^{(2)}(t,\tau)$ from the
stationary state ($t\to\infty$) of continuously and incoherently pumped
emitters for a) a single emitter and two distinguishable emitters 
(for $\gamma_p=\gamma$) and
b) two superradiant emitters with superradiant decay rate $\gamma_S=2\gamma$ and
different pump rates $\gamma_p$ without additional dephasing $\gamma_d=0$.
c) Superradiant emission from two emitters with additional local dephasing 
$\gamma_d$ at pump rate $\gamma_p=0.5\gamma$.}
\end{figure*}

Typically, photon coincidences are measured as a function of the delay time
$\tau$, which additionally includes information about the dynamics.
As long as the photon environment is not strongly structured as, e.g., in 
single-mode microcavities, and, hence, does not show significant non-Markovian 
memory effects, the time evolution can be well described by Lindblad 
master equations. 
For single, distinguishable, and superradiant emitters, which are
pumped incoherently, the respective master equations are
\begin{align}
\frac{\partial}{\partial t} \rho_\textnormal{single} = &
\gamma_p \mathcal{D}\big[\sigma^+_1\big]\big(\rho_\textnormal{single}\big)
+\gamma \mathcal{D}\big[\sigma^-_1\big]\big(\rho_\textnormal{single}\big),
\\
\label{eq:Lind_dist}
\frac{\partial}{\partial t} \rho_\textnormal{dist} = &
\gamma_p \mathcal{D}\big[\sigma^+_1\big]\big(\rho_\textnormal{dist}\big)
+\gamma_p \mathcal{D}\big[\sigma^+_2\big]\big(\rho_\textnormal{dist}\big)
\nn&+
\gamma \mathcal{D}\big[\sigma^-_1\big]\big(\rho_\textnormal{dist}\big)
+\gamma \mathcal{D}\big[\sigma^-_2\big]\big(\rho_\textnormal{dist}\big),
\\
\label{eq:Lind_sup}
\frac{\partial}{\partial t} \rho_\textnormal{sup} = &
\gamma_p \mathcal{D}\big[\sigma^+_1\big]\big(\rho_\textnormal{sup}\big)
+\gamma_p \mathcal{D}\big[\sigma^+_2\big]\big(\rho_\textnormal{sup}\big)
\nn&+
\gamma_d \mathcal{D}\big[\sigma^+_1\sigma^-_1\big]\big(\rho_\textnormal{sup}\big)
+\gamma_d \mathcal{D}\big[\sigma^+_2\sigma^-_2\big]\big(\rho_\textnormal{sup}\big)
\nn&+
\gamma_S \mathcal{D}\big[\sigma^-_S\big]\big(\rho_\textnormal{sup}\big),
\end{align}
respectively, where 
\begin{align}
\mathcal{D}\big[L\big]\big(\rho\big)=&
L\rho L^\dagger -\frac 12\big(L^\dagger L \rho + \rho L^\dagger L\big).
\end{align}
is the Lindblad superoperator, $\gamma$ is the radiative decay rate of a 
single emitter, $\gamma_p$ is the pump rate, and $\gamma_S=2\gamma$ is the
superradiant decay rate. Additionally, in the superradiant case, 
we have introduced local dephasing rates $\gamma_d$, 
which also leads to the decay of inter-emitter correlations.

Generally, two-time correlation functions of the form
$\langle a^\dagger_{\mathbf{k}}(t) a^\dagger_{\mathbf{k'}}(t+\tau)
a_{\mathbf{k}'}(t+\tau)a_{\mathbf{k}}(t)\rangle$ 
are obtained using the quantum regression theorem\cite{QRT_Lax,QRT}
by propagating a density matrix according to the respective 
Lindblad master equations up to time $t$. 
Applying the respective operators,
one defines the unnormalized pseudo density matrices
$\rho'(0)=a_{\mathbf{k}}\rho(t) a^\dagger_{\mathbf{k}}$. The latter is then
propagated using the same master equation for a time $\tau$, 
at which the correlation function is evaluated
as $\langle a^\dagger_{\mathbf{k}}(t) a^\dagger_{\mathbf{k'}}(t+\tau)$
$a_{\mathbf{k}'}(t+\tau)a_{\mathbf{k}}(t)\rangle = 
\textrm{Tr} \big[a^\dagger_{\mathbf{k'}} a_{\mathbf{k}'} \rho'(\tau)\big]$.

With this approach, the delay-time dependent coincidences
$g^{(2)}(t,\tau)$ from the stationary state $t\to \infty$ can be calculated
analytically for single and indistinguishable emitters as
(cf. Appendix \ref{app:single_dist})
\begin{align}
g^{(2)}_\textnormal{single}(\infty,\tau)=&
1 -e^{-(\gamma+\gamma_p)\tau},
\label{eq:g2single}
\\
g^{(2)}_\textnormal{dist}(\infty,\tau)=&1 - \frac 12 e^{-(\gamma+\gamma_p)\tau}.
\label{eq:g2dist}
\end{align}
The coincidences for the superradiant case are calculated numerically.

The delay-time dependence of $g^{(2)}(\infty,\tau)$ for the three cases are
depicted in Fig.~\ref{fig:dip}. As predicted analytically, 
the coincidences for single and distinguishable emitters 
show a dip at $\tau= 0$
with $g^{(2)}_\textnormal{single}(\infty,0)=0$ 
and $g^{(2)}_\textnormal{dist}(\infty,0)=1/2$, respectively.
In contrast, coincidences in the superradiant case feature an anti-dip 
with values $g^{(2)}_\textnormal{sup}(\infty,0) > 1/2$.
The height of the anti-dip $g^{(2)}_\textnormal{sup}(\infty,0)$ 
depends on the driving conditions like the pump rate $\gamma_p$ as shown in
Fig.~\ref{fig:dip}b.
In line with Eq.~\eqref{eq:g2sup0}, $g^{(2)}_\textnormal{sup}(t,0)=1$ 
when the emitters are uncorrelated at time $t$. This is the case for the
stationary state at special driving conditions 
$\gamma_p=\gamma=\gamma_S/2$.
Alternatively, the stationary state becomes uncorrelated if correlations 
introduced by driving and losses are suppressed by strong dephasing.
Indeed, as depicted in Fig.~\ref{fig:dip}c, 
$g^{(2)}_\textnormal{sup}(\infty,0)$ approaches 1 with increasing 
dephasing rate $\gamma_d$.

\section{Cooperative emission beyond superradiance}
So far, we have discussed distinguishable and superradiant emitters,
which are limiting cases of the Hamiltonian in Eq.~\eqref{eq:hamil}
where the phase factors $e^{\pm i\mathbf{k}\cdot\mathbf{r}}$ in the
light-matter coupling are either irrelevant or unity.
We now consider the more general regime of spectrally indistinguishable emitters
where the condition for free-space super\-radiance, namely inter-emitter distances
$r\ll\lambda$  being much smaller than wavelength of the light, 
is dropped. 
Then, the phase factors in the coupling play a crucial and non-trivial role. 
At the same time, the condition $\omega_1=\omega_2$ 
for spectrally indistinguishable emitters 
again precludes a straightforward application
of non-degenerate perturbation theory to describe the emission process.
Here, we solve this problem by describing the interaction with each 
light field mode labelled by its wave vector $\mathbf{k}$ as an 
independent decay channel.
For each channel, a situation analogous to that in the superradiant case 
emerges. The overall dynamics then follows from interference between
the individual decay processes. 

\subsection{Radiative decay}
\begin{figure}
\includegraphics[width=\linewidth]{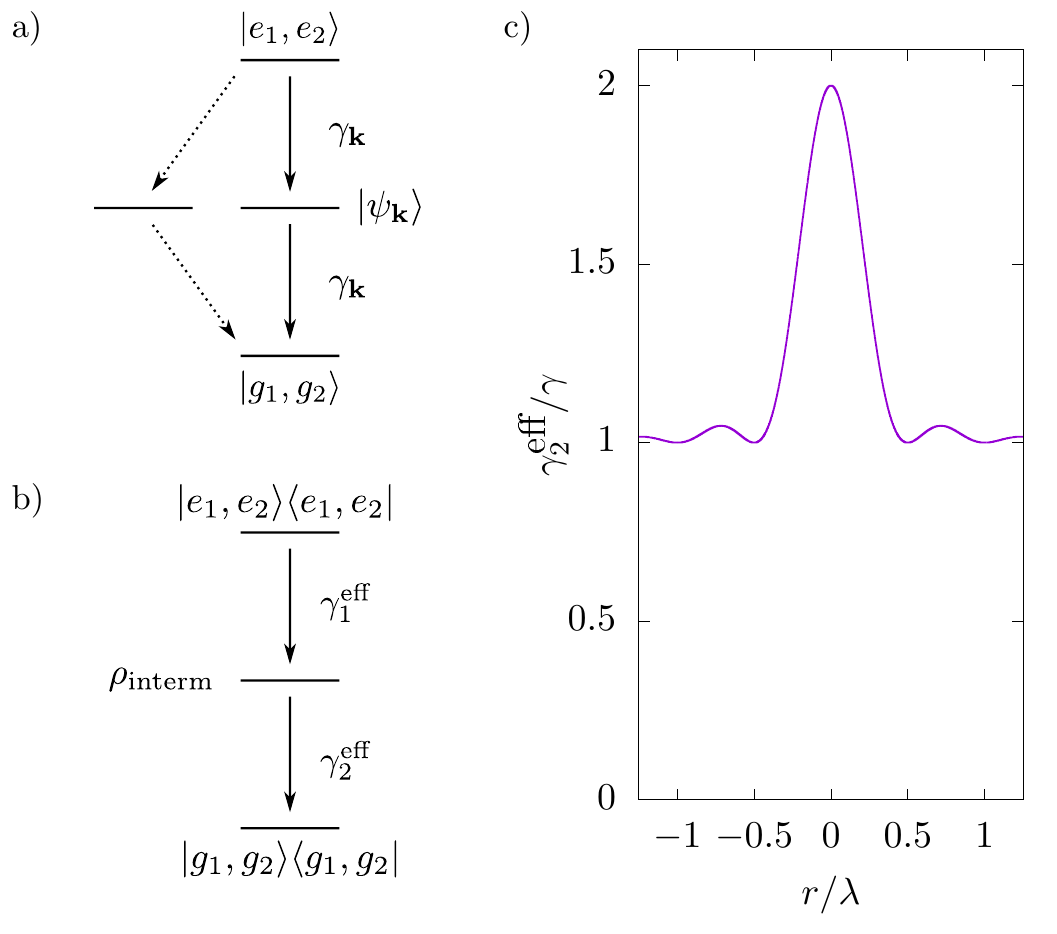}
\caption{\label{fig:sketch_k}a) Radiative decay processes for 
general indistinguishable emitters involving a single photon wave 
vector $\mathbf{k}$ and b) its overall effect using a density matrix
description. 
c) Effective emission rate $\gamma_2^\textnormal{eff}$
of the second photon as a function of the distance $r$ between the
emitters for detection angle $\theta_m\approx 54.74^\circ$.}
\end{figure}

In analogy to the superradiant case, we express the interaction Hamiltonian as
\begin{align}
H_I=&
\sum_{\mathbf{k}}\hbar\sqrt{2}g_\mathbf{k}\big(
\sigma^-_{\mathbf{k}}a^\dagger_{\mathbf{k}}
+\sigma^+_{\mathbf{k}}a_{\mathbf{k}}\big),
\end{align}
where we define the lowering and raising operators
\begin{align}
\sigma^-_{\mathbf{k}}=&\frac 1{\sqrt{2}}\big(
e^{i\mathbf{k}\cdot\mathbf{r}/2}\sigma^-_1
+e^{-i\mathbf{k}\cdot\mathbf{r}/2}\sigma^-_2\big)
\label{eq:defsigmak}
\end{align}
and $\sigma^+_{\mathbf{k}}=\big(\sigma^-_{\mathbf{k}}\big)^\dagger$,
which describe transitions 
\begin{align}
\sigma^-_{\mathbf{k}}=&
|g_1,g_2\rangle\langle\psi_{\mathbf{k}}|
+|\psi_{\mathbf{k}}\rangle\langle e_1,e_2|
\end{align}
through the intermediate state 
\begin{align}
|\psi_{\mathbf{k}}\rangle=& \frac 1{\sqrt{2}}\big(
e^{-i\mathbf{k}\cdot \mathbf{r}/2} |e_1,g_2\rangle
+ e^{i\mathbf{k}\cdot \mathbf{r}/2} |g_1,e_2\rangle \big).
\end{align}
Thus, for a fixed vector $\mathbf{k}$, the state $|\psi_{\mathbf{k}}\rangle$
plays a similar role as the symmetric Dicke state $|\psi_S\rangle$ in the
superradiant case (compare Fig.~\ref{fig:sketch_k}a with 
Fig.~\ref{fig:ind_sup}b),
albeit with different intermediate states 
$|\psi_{\mathbf{k}}\rangle$ and $|\psi_{\mathbf{k}'}\rangle$ for different wave
vectors $\mathbf{k}$ and $\mathbf{k}'$.
Each wave vector $\mathbf{k}$ constitutes a decay channel from the 
doubly excited state to the ground state via an intermediate state 
$|\psi_{\mathbf{k}}\rangle$, which is described by a Lindblad term
$\gamma_{\mathbf{k}}\mathcal{D}\big[\sigma^-_{\mathbf{k}}\big]\big(\rho\big)$
with rate  
(cf. Appendix~\ref{app:krate})
\begin{align}
\gamma_{\mathbf{k}}= {4\pi}{\hbar} g^2_\mathbf{k}
\delta(\hbar\omega_{\mathbf{k}}-\hbar\omega_i).
\end{align}

This description serves several purposes: On the one hand, 
the decay channels characterize the extraction of $\mathbf{k}$-dependent 
emitted intensities and, thus, links the radiation pattern to the 
quantum state of the emitter system. 
On the other hand, due to conservation of excitations, the free radiative decay
of the emitter system can be obtained by summing over the 
Lindbladians of all decay channels 
\begin{align}
\frac{\partial}{\partial t} \rho=& 
\sum_{\mathbf{k}} \gamma_{\mathbf{k}} 
\Big(\sigma^-_\mathbf{k}\rho\sigma^+_\mathbf{k} 
-\frac 12(\sigma^+_\mathbf{k}\sigma^-_\mathbf{k}\rho+	
\rho\sigma^+_\mathbf{k}\sigma^-_\mathbf{k})\Big).
\label{eq:Lind_sum}
\end{align}
With the concrete expressions for operators $\sigma^\pm_\mathbf{k}$ 
in Eq.~\eqref{eq:defsigmak}, we can alternatively write this master equation 
\begin{align}
\frac{\partial}{\partial t} \rho=& 
\gamma_\textrm{sup}\mathcal{D}\big[\sigma_S^-\big]\big(\rho\big)
+\gamma_\textrm{ind}\Big[\mathcal{D}\big[\sigma_1^-\big]\big(\rho\big)
+\mathcal{D}\big[\sigma_2^-\big]\big(\rho\big)\Big]
\label{eq:Lind_eff}
\end{align}
in terms of effective decay channels of the form of superradiant and
independent decay, respectively, with rates
\begin{align}
\gamma_\textrm{sup}=&
\sum_{\mathbf{k}} \gamma_{\mathbf{k}} e^{i\mathbf{k}\cdot\mathbf{r}},\\
\gamma_\textrm{ind}=&\sum_\mathbf{k}\gamma_{\mathbf{k}} 
\frac 12\big(1-e^{i\mathbf{k}\cdot\mathbf{r}}\big)
=\gamma-\frac 12\gamma_\textrm{sup},
\end{align}
where $\gamma$ is again the radiative decay rate of a single emitter.

Ignoring the radiation pattern of the dipole, i.e., assuming 
$\mathbf{k}$-independent couplings $g_\mathbf{k}=g$, integration over
$\mathbf{k}$ yields
\begin{align}
\gamma_\textrm{sup} =&2\gamma\frac{
\sum_{\mathbf{k}}\delta(\hbar\omega_\mathbf{k}-\hbar\omega) 
e^{i\mathbf{k}\cdot\mathbf{r}}}
{\sum_{\mathbf{k}}\delta(\hbar\omega_\mathbf{k}-\hbar\omega)} 
=2\gamma \frac{\sin(kr)}{kr}.
\label{eq:sinc}
\end{align}
Including dipole radiation of identical emitters 
by accounting for light polarization 
$\mathbf{e}_{\mathbf{k},\lambda}\perp \mathbf{k}$ 
perpendicular to the wave propagation via
$g_\mathbf{k,\lambda}=g(\hat{\mathbf{d}}\cdot\mathbf{e}_{\mathbf{k},\lambda})^2$,
where $\hat{\mathbf{d}}$ denotes the normalized direction of the dipoles,
and summing over $\lambda=1,2$ yields\cite{Haroche,Ficek}
\begin{align}
\gamma_\textrm{sup} =& 3 \gamma \bigg\{
\bigg[1-\frac{(\hat{\mathbf{d}}\cdot{\mathbf{r}})^2}{r^2}\bigg]
\frac{\sin kr}{kr} 
\nonumber\\&
+\bigg[1-3\frac{(\hat{\mathbf{d}}\cdot{\mathbf{r}})^2}{r^2}\bigg]
\bigg[\frac{\cos kr}{(kr)^2} -\frac{\sin kr}{(kr)^3}\bigg]\bigg\}.
\label{eq:sinc_dipole}
\end{align}
For simplicity, here, we focus on the situation of dipoles tilted by an 
angle $\theta_m\approx54.74^\circ$ with respect to the distance vector
between the emitters, where 
$(\hat{\mathbf{d}}\cdot{\mathbf{r}})^2/r^2=\cos^2\theta_m=1/3$, where
Eq.~\eqref{eq:sinc} is obtained as a special case of 
Eq.~\eqref{eq:sinc_dipole}. 

In this case, it is particularly clear that the 
master equation~\eqref{eq:Lind_eff} reproduces the radiative decay terms 
in Eqs.~\eqref{eq:Lind_dist} and~\eqref{eq:Lind_sup}
in the respective limits $r\gg \lambda$, 
where $\gamma_\textnormal{ind}=\gamma$ and $\gamma_\textnormal{sup}=0$,
and $r \ll \lambda$, where $\gamma_\textnormal{ind}=0$ and
$\gamma_S=\gamma_\textnormal{sup}=2\gamma$.

The net effect of the master equation~\eqref{eq:Lind_eff} for a system 
initially prepared in the doubly excited state $|e_1,e_2\rangle$ can be
visualized (cf. Fig.~\ref{fig:sketch_k}b) 
as transitions involving an intermediate state 
expressed as a mixed-state density matrix $\rho_\textnormal{interm}$.
This is obtained by equating the r.h.s. of Eq.~\eqref{eq:Lind_eff} 
evaluated for $\rho=|e_1,e_2\rangle\langle e_1,e_2|$ with
$\gamma_1^{\textnormal{eff}}\rho_\textnormal{interm}$, where 
$\rho_\textnormal{interm}$ is normalized to trace 1 and the norm 
defines the effective first photon emission rate 
$\gamma_1^{\textnormal{eff}}$.
This yields
\begin{align}
\rho_\textnormal{interm}=& \sum_\mathbf{k} \frac{\gamma_{\mathbf{k}}}{2\gamma}
|\psi_\mathbf{k}\rangle\langle\psi_\mathbf{k}|
\nonumber\\=&
\frac{1}{2}\Big[
|e_1,g_2\rangle\langle e_1,g_2| +
|g_1,e_2\rangle\langle g_1,e_2|
\nonumber\\&+
\frac{\sin(kr)}{kr}\big(
|e_1,g_2\rangle\langle g_1,e_2| +
|g_1,e_2\rangle\langle e_1,g_2|\big)  \Big].
\end{align}
The first photon emission rate $\gamma_1^{\textnormal{eff}}=2\gamma$,
irrespective of the distance between emitters.
The emission of a second photon from the intermediate state, on the other hand,
depends on the overlap of the intermediate state with the respective decay
channel
\begin{align}
\gamma_2^{\textnormal{eff}}=&\sum_\mathbf{k} \gamma_{\mathbf{k}}
\langle \psi_\mathbf{k}|\rho_{\textnormal{interm}}|\psi_\mathbf{k}\rangle
=\gamma \bigg[1+ \Big(\frac{\sin(kr)}{kr}\Big)^2\bigg].
\end{align}

The effective second photon emission rate $\gamma_2^{\textnormal{eff}}$ 
is shown in Fig.~\ref{fig:sketch_k}c as a function of the distance between the
emitters. For small distances $r\ll \lambda$, one recovers the superradiant
limit with $\gamma_2^{\textnormal{eff}}=\gamma_1^{\textnormal{eff}}=2\gamma$.
At finite distances, the second photon emission rate decreases and 
eventually reaches values around 
$\gamma_2^{\textnormal{eff}}=\gamma_1^{\textnormal{eff}}/2 = \gamma$ at
$r\gtrsim \lambda/2$.

Thus, for large distances, the free radiative decay of two indistinguishable
emitters behaves just like that of distinguishable or independent emitters, 
where two emitters can contribute to the first emission process,
while, once one emitter is deexcited, 
only the remaining emitter contributes to the emission of the second photon.
No sign of cooperative emission is found in this regime when 
only the free radiative decay is considered.

\subsection{Photon coincidences}
\begin{figure}
\includegraphics[width=\linewidth]{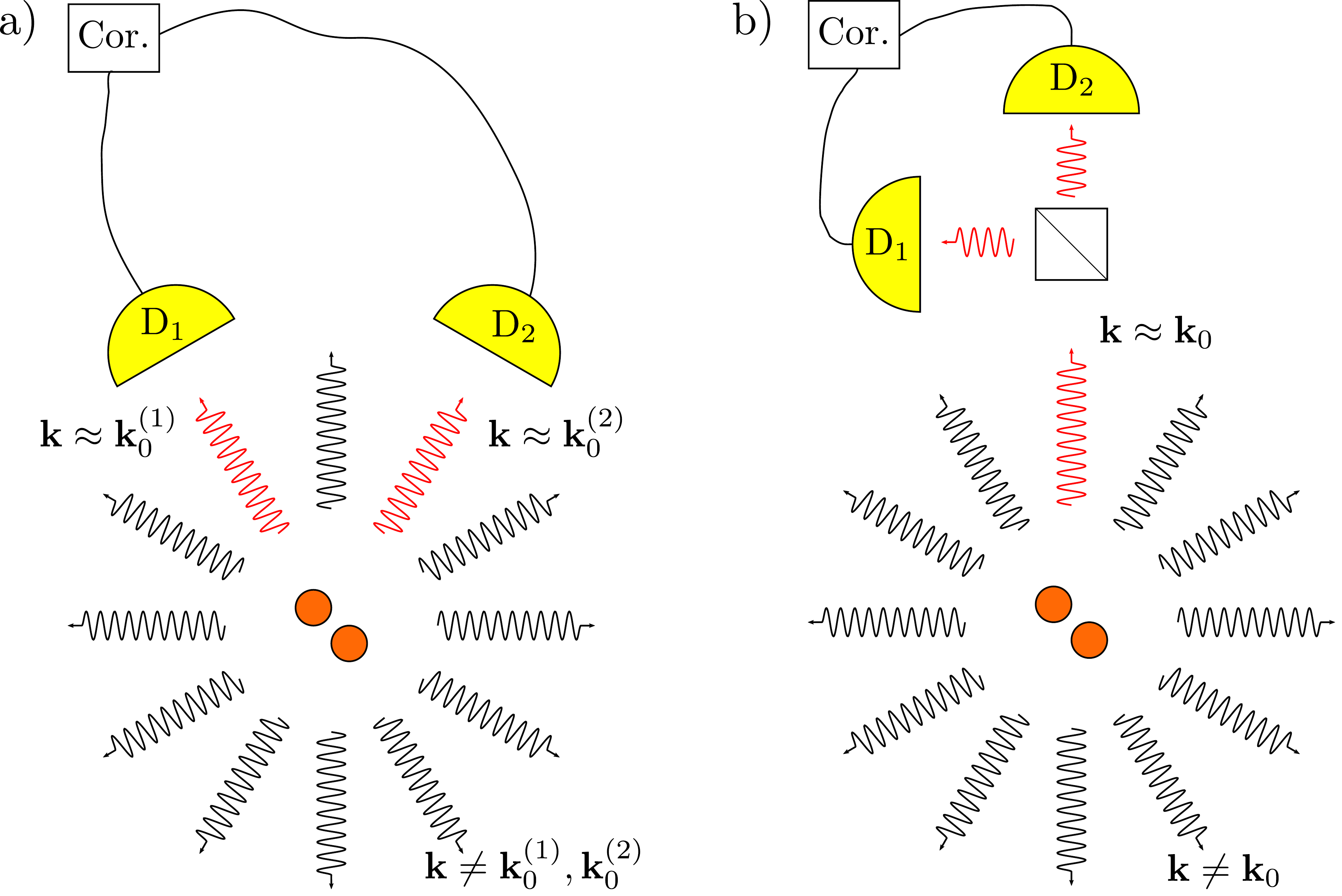}
\caption{\label{fig:detection}Sketch of photon coincidence setups. a) 
Coincidences recorded from two detectors. The detectors D$_1$ and D$_2$ 
predominantly register photons with wave vectors $\mathbf{k}$ close to 
reference wave vectors $\mathbf{k}_0^{(i)}$, which are determined by
the relative positioning of the detectors with respect to the emitters.
b) Hanbury Brown-Twiss setup measuring coincidences of photons in the same
direction $\mathbf{k}_0^{(1)}=\mathbf{k}_0^{(2)}=\mathbf{k}_0$.
Photons with different wave vectors 
$\mathbf{k}\neq \mathbf{k}_0^{(1)},\mathbf{k}_0^{(2)}$ 
are relevant for the dynamics as they take part in radiative decay processes, 
but do not contribute to the detected signals.
}
\end{figure}

The reason why radiative decay alone does not lead 
to visible cooperative emission effects is related to the integration over all
accessible light field modes (summation over $\mathbf{k}$).
This situation changes markedly when $\mathbf{k}$-resolved quantities
are considered\cite{Schmidt-Kaler2022,Schmidt-KalerPRL}.
For example, measuring photon coincidences using two detectors, 
as depicted in Fig.~\ref{fig:detection}, 
one usually collects photons emitted into a limited 
solid angle determined by the optical beam path as well as the position of
the detectors relative to the emitters.
Here, we assume two detectors with detection efficiencies 
$\eta_{\mathbf{k}}^{(1)}$ and $\eta_{\mathbf{k}}^{(2)}$ with finite support
narrowly localized around reference wave vectors $\mathbf{k}_0^{(1)}$ and
$\mathbf{k}_0^{(2)}$, respectively, i.e.,
\begin{align}
\eta_{\mathbf{k}}^{(l)}= \begin{cases}
\eta_0, &\mathbf{k}\approx\mathbf{k}_0^{(l)}, \\
0, &\textnormal{else}.
\end{cases}
\end{align}

The measured coincidence signal for wave-vector-selected photons can 
then be expressed as
\begin{align}
g^{(2)}_\textnormal{sel}(t,\tau)=& 
\frac{G^{(2)}_\textnormal{sel}(t,\tau)}
{I_\textnormal{sel}(t) I_\textnormal{sel}(t+\tau)},
\\
G^{(2)}_\textnormal{sel}(t,\tau)=&
\sum\limits_{\mathbf{k},\mathbf{k}'=\mathbf{k}_0^{(1)},\mathbf{k}_0^{(2)}}
 \langle \sigma^\dagger_{\mathbf{k}}(t)
\sigma^\dagger_{\mathbf{k}'}(t+\tau)
\sigma_{\mathbf{k}'}(t+\tau) \sigma_{\mathbf{k}}(t)\rangle,
\\
I_\textnormal{sel}(t)=&
\sum\limits_{\mathbf{k}=\mathbf{k}_0^{(1)},\mathbf{k}_0^{(2)}}
\langle \sigma^\dagger_{\mathbf{k}}(t)\sigma_{\mathbf{k}}(t)\rangle.
\end{align}

Because $\sigma_{\mathbf{k}_0^{(2)}}\sigma_{\mathbf{k}_0^{(1)}}=
|g_1,g_2\rangle\langle \psi_{\mathbf{k}_0^{(2)}}|
\psi_{\mathbf{k}_0^{(1)}}\rangle \langle e_1,e_2|$ 
with
$\langle \psi_{\mathbf{k}_0^{(2)}}|\psi_{\mathbf{k}_0^{(1)}}\rangle
=\cos\big[(\mathbf{k}_0^{(2)}-\mathbf{k}_0^{(1)})\cdot \mathbf{r}/2 \big]$,
the zero-delay coincidences become
\begin{align}
g^{(2)}_\textnormal{sel}(t, 0) =&  \bigg( \frac 34 +  \frac 14
\cos\big[(\mathbf{k}_0^{(2)}-\mathbf{k}_0^{(1)})\cdot\mathbf{r}\big]\bigg)
\nn&\times  
\frac{4 n_{e_1,e_2}(t)}
{\Big(2n_{e_1,e_2}(t)+n_{\mathbf{k}_0^{(1)}}(t)+n_{\mathbf{k}_0^{(2)}}(t)\Big)^2},
\label{eq:g2kk}
\end{align}
where $n_{\mathbf{k}}=\langle \sigma^\dagger_\mathbf{k}\sigma_\mathbf{k}\rangle$
are the occupations of the state $|\psi_\mathbf{k}\rangle$.
Note that the geometric factor 
$\big( 1+ \cos\big[(\mathbf{k}_0^{(2)}-\mathbf{k}_0^{(1)})\cdot\mathbf{r}\big]\big)$
can also be understood as interference of photons emitted from
the two emitters much like in a double-slit 
experiment~\cite{Walther,KeitelPRL,FicekBook}.

Specifically, for coincidence measurements with
$\mathbf{k}_0^{(2)}=\mathbf{k}_0^{(1)}=\mathbf{k}_0$ as obtained by 
a Hanbury Brown-Twiss (HBT) setup~\cite{HBT} as depicted in
Fig.~\ref{fig:detection}b), one finds
\begin{align}
\label{eq:g2sel}
g^{(2)}_{\textnormal{sel}}(t,0)=&
\frac{n_{e_1,e_2}(t)}{\big(n_{e_1,e_2}(t)+n_{\mathbf{k}_0}(t)\big)^2},
\end{align}
which is $g^{(2)}_{\textnormal{sel}}(t,0)=1$ 
for initially uncorrelated and equally occupied emitters.
Notably, this result, which coincides with that for superradiant emitters,
is independent of the distance $r$ between the dots and is therefore found 
even for $r\gtrsim \lambda$, where the free radiative decay shows no
superradiant enhancement.

This finding can be understood as follows: Having detected a single photon
from two indistinguishable emitters, no information is gained about the 
exact origin of the photon, i.e., whether it was emitted from emitter 1 
or emitter 2. Consequently, when the system is initially in the doubly excited
state, the detection event at detector~1 indicates a collapse of 
the wave function of the two-emitter system 
to the maximally entangled state $|\psi_{\mathbf{k}_0^{(1)}}\rangle$.
In contrast to the situation for an uncorrelated statistical mixture of
excitations in either emitter, emission of the second photon from the
correlated state $|\psi_{\mathbf{k}_0^{(1)}}\rangle$ 
has an oscillator strength that
strongly depends on the emission direction $\mathbf{k}_0^{(2)}$ 
of the second photon due to the overlap 
$\langle \psi_{\mathbf{k}_0^{(2)}}|\psi_{\mathbf{k}_0^{(1)}}\rangle
=\cos\big[(\mathbf{k}_0^{(2)}-\mathbf{k}_0^{(1)})\cdot \mathbf{r}/2 \big]$.
Thus, the emission rate for the second photon in the direction 
$\mathbf{k}_0^{(2)}=\mathbf{k}_0^{(1)}$ is enhanced without a
concomitant enhancement of the overall radiative decay rate 
by diverting oscillator strength from other emission directions 
$\mathbf{k}_0^{(2)}$, in particular from those for which
$(\mathbf{k}_0^{(2)}-\mathbf{k}_0^{(1)})\cdot\mathbf{r}$ is close to
an odd multiple of $\pi$, for which 
$g^{(2)}_\textnormal{sel}(t,0)\approx 0$.
This gives rise to a distinct radiation pattern for photon coincidences, 
which has been measured in Refs.~\cite{Schmidt-KalerPRL,Schmidt-Kaler2022}.

Finally, it is noteworthy that, although the distance $r$ between the emitters
is not relevant for the discussion of an ideal wave-vector-resolving 
HBT experiments ($\mathbf{k}_0^{(2)}=\mathbf{k}_0^{(1)}=\mathbf{k}_0$), 
there are practical limitations:
A realistic detector picks up wave vectors 
$\mathbf{k}=\mathbf{k}_0+\delta \mathbf{k}$ in a finite range 
$\delta \mathbf{k} \in \Omega$ 
around the reference wave vector $\mathbf{k}_0$.
Assuming that the detection efficiency $\eta_{\mathbf{k}}$ is constant
and non-zero only in $\Omega$, the registration of a photon at the detector
from an initially doubly excited two-emitter system can be described as
a collapse of the system to the mixed state

\begin{align}
\rho_{\textnormal{av}}= &
\frac 12\Big[
|e_1,g_2\rangle\langle e_1,g_2| +
|g_1,e_2\rangle\langle g_1,e_2|
\nn&+
\xi |e_1,g_2\rangle\langle g_1,e_2| +
\xi^* |g_1,e_2\rangle\langle e_1,g_2| \Big]
\end{align}
with
\begin{align}
\xi=& e^{-i\mathbf{k}_0\cdot\mathbf{r}}
\int\limits_{\Omega}\frac{d \delta\mathbf{k}}{\Omega}
e^{-i\delta\mathbf{k}\cdot\mathbf{r}}.
\end{align}
Consequently, the photon coincidences at zero delay for initially uncorrelated 
and equally occupied emitters are found to be
\begin{align}
g^{(2)}_{\textnormal{av}} (t, 0)=& \frac{1+|\xi|^2}2. 
\label{eq:g2av}
\end{align}

$|\xi|^2$ can be regarded as a measure for the precision of the wave vector
selection. As such, it determines how well cooperative effects 
of the emission from indistinguishable emitters are resolved by the detectors.
Ideal detectors resolving single wave vectors $\mathbf{k}_0$ in a 
HBT setup produce $|\xi|^2=1$ and therefore $g^{(2)}(t,0)=1$, whereas 
for detectors with finite detection ranges $\Omega$,
$|\xi|^2$ is reduced due to cancellation, which becomes significant for
large distances $r\gtrsim 2\pi /\delta {k}_\textnormal{max}$, 
where $\delta k_\textnormal{max}$ is of the order of
the maximal spread of wave vectors within the detectable range $\Omega$. 

Thus, for realistic detectors and for large distances between the emitters, 
$\xi$ approaches 0 leading to zero-delay photon coincidence  
$g^{(2)}(t,0)=1/2$, similar to that for distinguishable emitters.
This sets a natural limitation for the detection of cooperative effects,
which is, however, much less stringent that the condition
$r<\lambda$ for superradiance.

\subsection{Delay-time dependence of photon coincidences}

We now consider the delay-time dependence of photon coincidences
in the case of an ideal HBT setup ($\mathbf{k}_0^{(2)}=\mathbf{k}_0^{(1)}
=\mathbf{k}_0$, $|\xi|^2=1$) 
in the non-superradiant regime 
$r\gtrsim \lambda$ of cooperative emission by selective measurement. 
We assume continuously and incoherently driven identical emitters subject to 
radiative decay and additional dephasing with rate $\gamma_d$ 
(e.g., due to interactions with a phonon bath), whose free evolution is
described by the Lindblad master equation
\begin{align}
\frac{\partial}{\partial t}\rho_\textnormal{sel}=& 
\gamma_p\mathcal{D}\big[\sigma_1^+\big]\big(\rho_\textnormal{sel}\big)
+\gamma_p\mathcal{D}\big[\sigma_2^+\big]\big(\rho_\textnormal{sel}\big)
\nn&
+\gamma_d\mathcal{D}\big[\sigma_1^+\sigma_1^-\big]\big(\rho_\textnormal{sel}\big)
+\gamma_d\mathcal{D}\big[\sigma_2^+\sigma_2^-\big]\big(\rho_\textnormal{sel}\big)
\nn&
+\gamma\mathcal{D}\big[\sigma_1^-\big]\big(\rho_\textnormal{sel}\big)
+\gamma\mathcal{D}\big[\sigma_2^-\big]\big(\rho_\textnormal{sel}\big).
\label{eq:Lind_indist}
\end{align}

\begin{figure*}
\includegraphics[width=\linewidth]{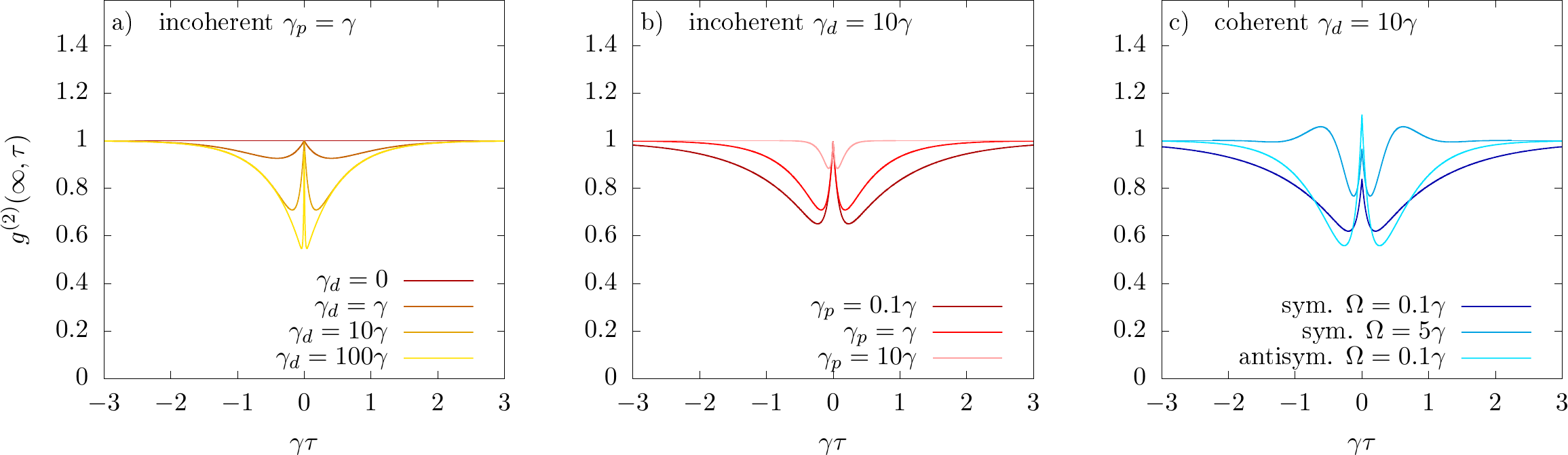}
\caption{\label{fig:coop}Delay-time dependence of photon coincidences 
$g^{(2)}(t\to\infty,\tau)$ for cooperative emission due to 
selective measurement.
a) Coincidences for incoherent pumping with rate $\gamma_p=\gamma$ and for
different dephasing rates $\gamma_d$.
b) Incoherent pumping with different rates $\gamma_p$ at fixed dephasing 
rate $\gamma_d=10\gamma$.
c) Coincidences for coherently driven indistinguishable emitters with 
dephasing rate $\gamma_d=10\gamma$ for different Rabi frequencies $\Omega$. 
The driving either couples to both emitters with the same phase (``sym.'') or 
with opposite phases (``antisym'').
}
\end{figure*}
The analysis can be simplified by considering identical emitters
with identical occupations
$n_{e_1,g_2}=\langle e_1,g_2|\rho|e_1,g_2\rangle=
\langle g_1,e_2|\rho|g_1,e_2\rangle=n_{g_1,e_2}$ and 
choosing $e^{i\mathbf{k}_0\cdot\mathbf{r}}=1$. Note that, 
for fixed $\mathbf{k}_0$, a nonzero phase can be eliminated 
by redefining the phase of state $|e_2\rangle$.
With this choice of phase, the only non-zero off-diagonal element of the
4-level density matrix, the inter-emitter coherences  
$c=\langle e_1,g_2| \rho | g_1,e_2\rangle$, remain real $c^*=c$ and
the intermediate state $|\psi_{\mathbf{k}_0}\rangle$ coincides with the
symmetric Dicke state $|\psi_S\rangle$.
Furthermore, the trace condition $\textrm{Tr}(\rho)=1$ is used 
to express the ground state population as 
$n_{g_1,g_2}=1-n_{e_1,e_2}-2n_{e_1,g_2}$.
Noting further that the Lindblad master equation~\eqref{eq:Lind_indist}
never introduces coherences between states with different numbers of 
excitations, the dynamics can be fully described by the degrees of freedom 
$n_{e_1,e_2}, n_{e_1,g_2},$ and $c$, which obey the equation of motion
\begin{align}
&\frac{\partial}{\partial t}
\left(\begin{array}{c} n_{e_1,e_2} \\ n_{e_1,g_2} \\ c \end{array}\right)
= A \left(\begin{array}{c} n_{e_1,e_2} \\ n_{e_1,g_2} \\ c \end{array}\right)
+ \left(\begin{array}{c} 0 \\ \gamma_p \\ 0 \end{array}\right),
\\
&A=\left(\begin{array}{ccc}
-2 \gamma  & 2 \gamma_p & 0 \\
(\gamma- \gamma_p) & -\gamma -3\gamma_p  & 0 \\
0 & 0 & -\gamma -\gamma_p - \gamma_d
\end{array}\right).
\end{align}

First of all, we see that the coherences decouple from the dynamics of the
remaining degrees of freedom and initial coherence $c(0)$, e.g., introduced
by the measurement process, simply decay according to
\begin{align}
c(t)=&c(0)e^{-(\gamma+\gamma_p+\gamma_d)t}.
\end{align}
The remaining two-dimensional linear inhomogeneous ordinary differential 
equation can, in principle, be solved analytically. Especially compact 
results are found in the special case of equal pumping and decay
$\gamma_p=\gamma$, where the singly excited state population $n_{e_1,g_2}$
is decoupled from the doubly excited state population $n_{e_1,e_2}$. 
The corresponding equation
\begin{align}
\frac{\partial}{\partial t} n_{e_1,g_2}=& -4\gamma n_{e_1,g_2} + \gamma
\end{align}
is solved by
\begin{align}
n_{e_1,g_2}(t)=& \frac 14 +\Big(n_{e_1,g_2}(0) -\frac 14\Big)e^{-4\gamma t}.
\end{align}
This expression acts as a driving term in the equation for the doubly
excited state populations
\begin{align}
\frac{\partial}{\partial t} n_{e_1,e_2}=&
-2\gamma n_{e_1,e_2} +2 \gamma  n_{e_1,g_2},
\end{align}
which is solved by 
\begin{align}
n_{e_1,e_2}(t)=&
\frac 1{4}
+\Big(n_{e_1,e_2}(0) + n_{e_1,g_2}(0)-\frac 1{2} \Big) e^{-2\gamma t}
\nn&
+\Big(\frac 14 - n_{e_1,g_2}(0) \Big) e^{-4\gamma t}.
\end{align}

Due to our choice of phase $e^{i\mathbf{k}_0\cdot\mathbf{r}}=1$, 
the optical signals are described by
\begin{align}
\label{eq:G2sel}
G^{(2)}_\textnormal{sel}(t,\tau)=& 4I_0^2 
\langle \sigma_S^+(t)\sigma_S^+(t+\tau)\sigma_S^-(t+\tau)\sigma_S^-(t)\rangle,
\\
I_\textnormal{sel}(t)=& 2I_0 \langle \sigma_S^+(t)\sigma_S^-(t)\rangle 
\nn=&
2I_0 \big(n_{e_1,e_2}(t)+n_{e_1,g_2}(t)+c(t)\big), 
\end{align}
where we have used $n_S=\frac 12( n_{e_1,g_2}+n_{g_1,e_2}+c+c^*)=n_{e_1,g_2}+c$.

To obtain the normalized coincidences $g^{(2)}_\textnormal{sel}(t,\tau)$
for emission from the stationary state at $t\to\infty$,
\mbox{$G^{(2)}_\textnormal{sel}(t\to\infty,\tau)$} and 
$I_\textnormal{sel}(t\to \infty)$ can be analyzed individually. 
First, observing that the Lindblad master equation~\eqref{eq:Lind_indist}
reduces coherences, one finds \mbox{$c(t\to \infty)\to 0$}, 
so the stationary intensity
$I_\textnormal{sel}=2I_0 (n_{e_1,e_2}+n_{e_1,g_2})
=2 I_0 n_{e_1}$ becomes identical to $I_\textnormal{dist}$ for
two distinguishable emitters.

Applying operators $\sigma_S^\pm$ at time $t$, one finds two
nonzero contributions to the unnormalized coincidences 
$G^{(2)}_\textnormal{sel}(t,\tau)$ in Eq.~\eqref{eq:G2sel}:
i) With probability $2I_0n_{e_1,e_2}(t)$, a photon originating from the
doubly excited state is detected. This measurement process leads to the
collapse of the wave function to the state $|\psi_S\rangle$, which serves as 
the initial state for the delay-time propagation with 
$n'_{e_1,e_2}(\tau=0)=0$ and
$n'_{e_1,g_2}(\tau=0)=c'(\tau=0)=1/2$, where $n'_{e_1,e_2}$, 
$n'_{e_1,g_2}$, and $c'$ refer
to the occupations and coherences of the pseudo density matrix $\rho'$ 
in the quantum regression theorem.
ii) With probability $2I_0 n_S(t)$, a photon originating from the 
single-excitation manifold is detected and the system collapses onto the 
ground state implying $n'_{e_1,e_2}(\tau=0)=n'_{e_1,g_2}(\tau=0)=c'(\tau=0)=0$
for the delay-time propagation. In total, we find for $\gamma_p=\gamma$
\begin{align}
\label{eq:g2coop}
g^{(2)}_\textnormal{sel}(\infty,\tau)=& 
1-\frac 12 e^{-2\gamma \tau} 
+\underbrace{\frac 12|\xi|^2 e^{-(\gamma+\gamma_p+\gamma_d)\tau}}_{c'(\tau)},
\end{align}
where we have re-introduced the factor $|\xi|^2\le 1$ accounting for
the finite range of wave vectors detected by a realistic detector.

It is noteworthy that the first two terms in Eq.~\eqref{eq:g2coop} 
are identical to Eq.~\eqref{eq:g2dist} for distinguishable emitters, 
which leads to a dip at $\tau\to 0$ with 
$g^{(2)}_\textnormal{dist}(\infty,0)=1/2$. 
Here, however, the last term
in Eq.~\eqref{eq:g2coop} provides an additive contribution that directly 
reflects the measurement-induced correlations $c'(\tau)$ brought about by 
the collapse of the wave function due to photon detection 
from the doubly excited state and leaving the emitters in the 
correlated states $|\psi_S\rangle$. 
Perfect state preparation $|\xi|^2=1$, thus, leads to an anti-dip 
at $\tau\to 0$ with $g^{(2)}_\textnormal{sel}(\infty,0)=1$ 
and with a width determined by a combination of pump, decay, and dephasing 
rates.

While Eq.~\eqref{eq:g2coop} is derived for the special 
driving condition of equal pump and decay rates $\gamma_p=\gamma$,
a more comprehensive picture is gained by solving Eq.~\eqref{eq:Lind_indist}
numerically. 
In Fig.~\ref{fig:coop}a and b, photon coincidences for the situation 
of cooperative emission due to selective measurement 
are depicted for different values of dephasing rates $\gamma_d$ 
(for $\gamma_p=\gamma$) and pumping rates $\gamma_p$ 
(for $\gamma_d=10\gamma$), respectively.
In the absence of dephasing $\gamma_d=0$, coincidences remain 
$g^{(2)}(\infty,\tau)=1$ for all delay times $\tau$, which also 
follows from the cancellation of 
the second and the last terms in Eq.~\eqref{eq:g2coop}.
For finite dephasing $\gamma_d\neq 0$, the cancellation is incomplete, 
effectively leading to an anti-dip similar to that observed in the superradiant
case.
The width of the anti-dip decreases with increasing dephasing rate.
Increasing the pump rate $\gamma_p$ leads to a faster recovery of the
stationary coincidences with values $g^{(2)}(\infty,\tau)=1$.

It is interesting to note that the zero-delay coincidences in
Fig.~\ref{fig:coop}a and b remain $g^{(2)}(\infty,0)=1$ for all dephasing
rates $\gamma_d$ and pump rates $\gamma_p$.
This is due to the fact that the zero-delay coincidences are fully determined
by the stationary state and the measurement operators and do not probe
dynamical aspects. Here, the stationary state is uncorrelated as
the Lindblad master equation~\eqref{eq:Lind_indist} 
contains no term introducing correlations. According to Eq.~\eqref{eq:g2sel},
this entails unity zero-delay coincidences. 
In contrast, the master equation~\eqref{eq:Lind_sup} for the superradiant case
includes the superradiant decay described by the operator $\sigma^-_S$, 
which does lead to correlations in the stationary state, hence 
deviations from $g^{(2)}(\infty,0)=1$ were found in Fig.~\ref{fig:dip}.

However, non-unity values of the zero-delay coincidences may be obtained
also in the case of cooperative emission in the absence of superradiance, 
as long as
correlations in the stationary state are introduced by other means, e.g., 
by coherent driving of the two-emitter system.
To demonstrate this, we present in
Fig.~\ref{fig:coop}c results for the coincidences obtained when 
the incoherent pumping is replaced by coherent driving described by the
Hamiltonian 
\begin{align}
H_\textnormal{sym} =& \frac{\hbar}2 \Omega 
\big[( \sigma^+_1 + \sigma^-_1) + ( \sigma^+_2 + \sigma^-_2)\big]
\end{align}
for driving of emitters with equal phases and
\begin{align}
H_\textnormal{antisym} =& \frac{\hbar}2 \Omega 
\big[( \sigma^+_1 + \sigma^-_1) - ( \sigma^+_2 + \sigma^-_2)\big]
\end{align}
for driving with opposite phases.
\begin{figure*}
\includegraphics[width=\textwidth]{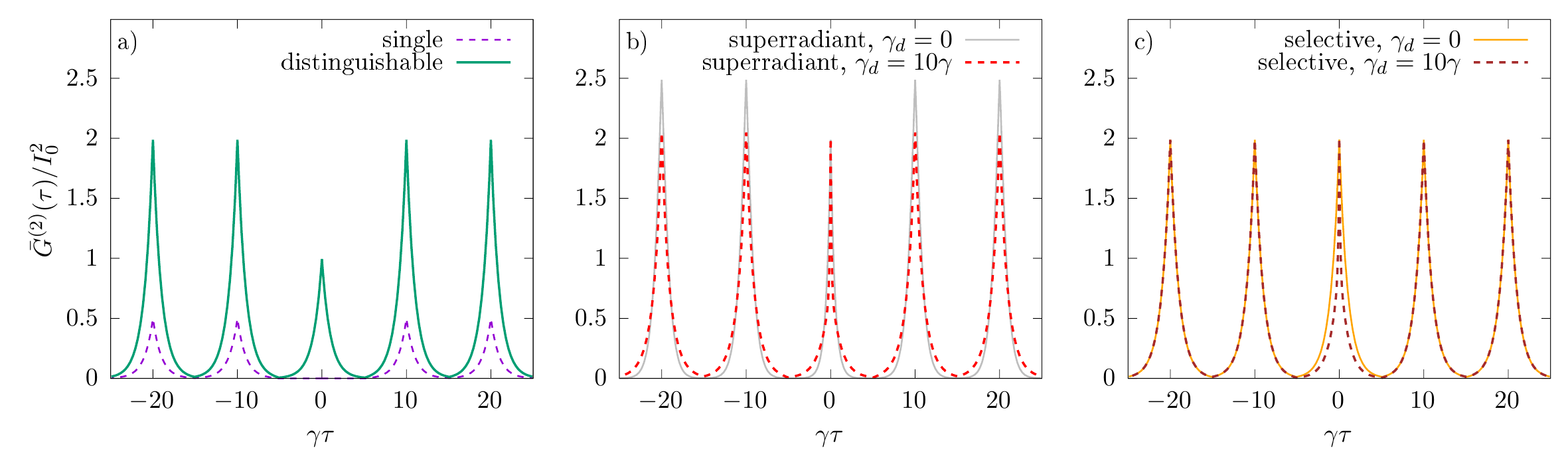}
\caption{\label{fig:pulsed}Delay-time dependence of photon coincidences 
$\bar{G}^{(2)}(\tau)$ after delta-like excitation with a pulse train
with repetition time $T=10/\gamma$ for
a) single and two distinguishable emitters,
b) superradiant emitters without $\gamma_d=0$ and with dephasing 
$\gamma_d=10\gamma$, and
c) cooperative emission due to selective measurement.
}
\end{figure*}

For weak coherent driving with equal phases, we find zero-delay coincidences
$g^{(2)}_\textnormal{sel}(\infty, 0)<1$. Increasing 
the driving strength eventually leads to shoulders at finite delay times 
indicative of Rabi oscillations, yet the zero-delay coincidences remain 
suppressed compared to incoherent pumping. 
However, weak driving with opposite phases indeed results in 
zero-delay coincidences exceeding one.
This can again be explained by Eq.~\eqref{eq:g2sel}, which predicts that 
large values of coincidences are favoured by settings 
where the stationary occupations $n_{e_1,e_2}\ll 1$ while 
still $n_S \ll n_{e_1,e_2}$. Such a situation can be achieved by weakly
driving the doubly excited state via the antisymmetric Dicke state 
$|\psi_A\rangle$.

\subsection{Photon coincidences from pulsed driving}
Beside continuous driving conditions, photon coincidences are also
frequently investigated using pulsed laser 
excitation\cite{Clark2022,Otten2020}, e.g., to assess the quality of
on-demand single photon sources\cite{Kiraz,Thomas2021,PI_singlephoton} 
or for quantum state tomography of sources of polarization-entangled 
photon pairs\cite{Stevenson08,ConcurrenceCygorek}.
In contrast to continuous driving, photon coincidences $G^{(2)}(t,\tau)$ 
under pulsed excitation depend explicitly on the detection time $t$ 
of the first photon. 
While suitably time-integrated and normalized coincidences agree
with the zero-delay coindences obtained under continuous driving in the
special cases of single and distinguishable emitters (see below), 
care has to be taken when interpreting more complex situations,
where the results may differ drastically depending, e.g., on the 
choice of time integration windows\cite{Stevenson08,ConcurrenceCygorek}.
Therefore, we now derive photon coincidences under pulsed excitation
for superradiant and cooperatively emitting emitters.

We consider emitters excited by a train of 
delta-like $\pi$-pulses with repetition time $T$. This time is chosen 
to be $T\gg 1/\gamma$ larger than the radiative decay time so that all 
excitations induced by one pulse have decayed before the next pulse
arrives. 
Typically, one integrates over the arrival time $t$ of the first photon.
The recorded histogram as a function of the delay time is proportional  
\begin{align}
\bar{G}^{(2)}(\tau):= \int\limits_{-T/2}^{T/2} dt\, G^{(2)}(t,\tau),
\end{align}
where we extend the domain of $\tau$ to negative delay times by setting
$\bar{G}^{(2)}(-\tau)=\bar{G}^{(2)}(\tau)$.

The signal $\bar{G}^{(2)}(\tau)$ has the form of 
a series of distinct peaks as a function of the delay time $\tau$ as
depicted in Fig.~\ref{fig:pulsed}.
The peak around $\tau= 0$ originates from excitations 
during a single pulse. The peaks around $\tau = n T$ with integer
$n\neq 0$ are due to excitations generated from different pulses.
Hence, the peaks at $n\neq 0$ correspond to independent emission events
and can be used as a normalization reference for photon coincidences 
of the zero-order peak.

There are two common ways to extract concrete figures of merit from the
delay-time-dependent function $\bar{G}^{(2)}(\tau)$ given by i) the heights 
and ii) the integrals of the respective peaks. Both can be defined as
\begin{align}
\label{eq:defbarg}
\bar{g}^{(2)}_{\Delta \tau}:=&\frac{\bar{G}^{(2)}_{0,\Delta \tau}}{
\bar{G}^{(2)}_{1,\Delta \tau}}. 
\end{align}
in terms of the delay-time integral 
\begin{align}
\bar{G}^{(2)}_{n,\Delta \tau}:=& 
\int\limits_{nT-\Delta \tau/2}^{nT+\Delta \tau/2}d\tau\, \bar{G}^{(2)}(\tau),
\end{align}
with delay-time integration window of width $\Delta \tau$.
The peak heights (i) are obtained in the limit $\Delta \tau\to 0$, whereas the
integrals over peaks (ii) are obtained for $\Delta \tau\to T$. 

First, we focus on the calculation of $\bar{G}^{(2)}_{1,\Delta \tau}$ for
peak $n=1$. The assumption $T\gg 1/\gamma$ implies that the first and second
detected photons originate from different pulses and the emitters had 
relaxed to the ground state in the time between emission events. Thus,
the states of the system at times $t$ and $t+\tau$ are uncorrelated, so 
correlation functions 
$\langle \sigma^+_i(t)\sigma^+_{j}(t+\tau)\sigma^-_{k}(t+\tau)\sigma^-_l(t)\rangle$
factorize into the product
$\langle \sigma^+_i(t)\sigma^-_l(t)\rangle
\langle\sigma^+_{j}(t+\tau)\sigma^-_{k}(t+\tau)\rangle$.
Consequently,
\begin{align}
G^{(2)}(t,t+\tau)=I(t)I(t+\tau),\quad \tau\gtrsim T
\end{align}
and one obtains
\begin{align}
\label{eq:G2_10}
\bar{G}^{(2)}_{1,\Delta \tau\to 0}=&
\Delta \tau  \int\limits_{0}^{T/2}dt\, I^2(t), \\
\label{eq:G2_1T}
\bar{G}^{(2)}_{1,\Delta \tau\to T}=&
\bigg[\int\limits_{0}^{T/2}dt\, I(t)\bigg]^2,
\end{align}
for short and long integration windows, respectively.

For the zeroth-order peak, the correlation functions have to be caculated
explicitly as
\begin{align}
\bar{G}^{(2)}_{0,\Delta \tau\to 0}=& 
\Delta \tau \int\limits_{0}^{T/2} dt\, G^{(2)}(t,0) \\
\bar{G}^{(2)}_{0,\Delta \tau\to T}=& 
2 \int\limits_{0}^{T/2} dt \int\limits_0^{T/2}d\tau\, G^{(2)}(t,\tau).
\end{align}

In the case of a single emitter (cf. Fig.~\ref{fig:pulsed}a), 
the assumption of delta-like short pulses
makes re-excitation impossible, so that at all times at most one excitation is 
present in the system. Hence, $G^{(2)}_\textnormal{single}(t,\tau)=0$ and also 
$\bar{g}^{(2)}_{\Delta \tau,\textnormal{single}}=0$, irrespective of the 
integration window and other details. 

For two distinguishable emitters which are only subject to individual radiative
decay (cf. Fig.~\ref{fig:pulsed}a),
\begin{align} 
I_\textnormal{dist}(t)=&2I_0 n(0) e^{-\gamma t},  \\
G^{(2)}_\textnormal{dist}(t,\tau)=&
2 I_0^2 n^2(0) e^{-\gamma t} e^{-\gamma(t+\tau)}, 
\end{align}
where $n(0)$ is the excited state occupation per emitter immediately after the
pulse. This directly results in
\begin{align}
\bar{g}^{(2)}_{\Delta \tau,\textnormal{dist}}= \frac 12
\end{align}
irrespective of the integration window $\Delta \tau$. 
The fact that this value
coincides with $g^{(2)}_\textnormal{dist}(\infty,0)$ for continuously 
driven emitters is a direct consequence of the independent dynamics of both 
emitters and the effects of time-averaging being equal 
for numerator and denominator of Eq.~\eqref{eq:defbarg}, because the 
time-dependence of both is given by an exponential decay with the same rate.

However, the similarity between $\bar{g}^{(2)}_{\Delta \tau}$ and 
the zero-delay coincidences under continuous driving $g^{(2)}(\infty,0)$
does not carry over to cases with more complex dynamics. 
In Appendix~\ref{app:pulsed_sup}, we derive the integrated coincidences 
for two superradiant emitters in absence of additional dephasing
\begin{align}
\bar{g}^{(2)}_{\Delta\tau\to 0,\textnormal{sup}} =& 
\frac{n_{e_1,e_2}(0)}{\frac 54 n_{e_1,e_2}^2(0) +\frac 12 n_{S}^2(0)
+\frac 32 n_{e_1,e_2}(0) n_S(0)},
\\
\bar{g}^{(2)}_{\Delta\tau \to T,\textnormal{sup}} =&
\frac{2n_{e_1,e_2}(0)}{\big(2n_{e_1,e_2}(0)+n_S(0)\big)^2}.
\end{align}

Here, the values for different integration windows differ and are also not
comparable to $g^{(2)}_\textnormal{sup}(\infty, 0)$ for the continuously driven
superradiant system derived in Eq.~\eqref{eq:g2sup0}. 
This is primarily due to the non-exponential dynamics of the symmetric Dicke 
state occupations, which results in different time-averaging effects
in numerator and denominator of $\bar{g}^{(2)}$.
Note also that the results depend on the exact values of occupations
of doubly excited and symmetric Dicke states created by the driving pulse.
For example, if the doubly excited state is fully occupied, one finds
$\bar{g}^{(2)}_{\Delta\tau\to 0,\textnormal{sup}}=\frac 45$ and
$\bar{g}^{(2)}_{\Delta\tau \to T,\textnormal{sup}}=\frac 12$. The latter value
is the same as for distinguishable emitters. 
Thus, $\bar{g}^{(2)}_{\Delta\tau \to T}$ fails to clearly indicate 
cooperative emission even for ideal superradiance.

In Fig.~\ref{fig:pulsed}b, we present numerical calculations of $\bar{G}(\tau)$
in the superradiant case obtained using 
the master equations~\eqref{eq:Lind_sup} without and with dephasing with rate
$\gamma_d$, where, instead of pumping ($\gamma_p=0$), the state of the emitter
is periodically reset to the doubly excited state with repetition time 
$T=10/\gamma$. Indeed, the ratio between the heights of the zeroth-order versus
first-order peaks is
$\bar{g}^{(2)}_{\Delta\tau\to 0,\textnormal{sup}}=1.25$ in the absence
of dephasing. We have checked that the ratio between the integrals of the
first and second peaks also reproduces the analytical value of 
$\bar{g}^{(2)}_{\Delta\tau\to T,\textnormal{sup}}=0.5$.
It is noteworthy that numerical simulations with dephasing $\gamma_d=10\gamma$
reveal an increase of $\bar{g}^{(2)}_{\Delta\tau\to 0,\textnormal{sup}}$ 
extracted from the peak heights to a value of about $1.03$, 
while the ratio between the integrals remains the same
$\bar{g}^{(2)}_{\Delta\tau\to T,\textnormal{sup}}=0.5$.

The time-integrated coincidences for the case of cooperative emission 
by selective measurement are derived in Appendix~\ref{app:pulsed_sel}.
If the emitters are uncorrelated immediately after excitation, these reduce
to
\begin{align}
\bar{g}^{(2)}_{\Delta\tau\to 0,\textnormal{sel}} =& 1 ,\\
\label{eq:bar_g2_T_sel}
\bar{g}^{(2)}_{\Delta\tau\to T,\textnormal{sel}} =& \frac 12
\bigg[1+\frac{\gamma}{\gamma+\gamma_d}\bigg].
\end{align}
Here, we find that photon coincidences after pulsed excitation with short
delay-time integration windows lead to similar results
as $g^{(2)}_{\textnormal{sel}}(\infty,0)$ obtained under continuous driving.
For wider windows, dephasing eventually reduces
the integrated intensities, bringing it closer to the value of $\frac 12$ 
for independent emitters in the limit $\gamma_d\to\infty$.
Numerically simulated $\bar{G}(\tau)$ depicted in Fig.~\ref{fig:pulsed}c 
show that, in the absence of dephasing, the zeroth-order peak is identical
to the first-order peak, hence 
$\bar{g}^{(2)}_{\Delta\tau\to 0,\textnormal{sel}}=
\bar{g}^{(2)}_{\Delta\tau\to T,\textnormal{sel}}=1$. 
Dephasing with rate $\gamma_d=10\gamma$ 
only narrows the width of the zeroth-order peak, resulting in 
the same heights $\bar{g}^{(2)}_{\Delta\tau\to 0,\textnormal{sel}}=1$
but smaller integrals 
$\bar{g}^{(2)}_{\Delta\tau\to T,\textnormal{sel}}=\frac{6}{11}\approx 0.55<1$. 

\section{Discussion}
We have investigated cooperative emission from two two-level quantum emitters,
where both emitters contribute to both photon emission events.
We compared two sources of cooperativity, superradiance and the preparation 
of correlated states by emission-angle-selective measurement.
Superradiance and measurement-induced cooperativity
require the emitters to be spectrally indistinguishable, but the former
has stricter requirements in terms of spatial indistinguishabilty:
A superradiant enhancement of the radiative decay rate necessitates that
the electromagnetic environment cannot distinguish between emission 
from either dot. This can be achieved by confining emitters into regions 
much smaller than the wavelength of the emitted light\cite{Dicke, Haroche}
or by exploiting photonic structures like waveguides\cite{Waks2018,Gammon2019}.
In contrast, measurement-induced cooperativity merely 
requires the detectors not to differentiate between photons from either dot,
which depends more on the optical beam path than on the concrete spatial
separation and, hence, can be found even when emitters in free space 
are separated by distances $r\gtrsim\lambda$ larger than the wavelength 
of the light.

The most tangible difference between the two forms of cooperative emission is
that, in the case of superradiance, the radiative decay rate is enhanced, 
whereas the overall free decay rate after measurement-induced cooperativity
remains the same as for independent emitters. However,
the induced correlations shape the 
radiation pattern\cite{Schmidt-KalerPRL,Schmidt-Kaler2022}, so that
a second photon is more likely funnelled into the direction of
the detectors at the expense of other emission directions.
This difference makes both cases distinguishable via time-resolving 
measurements of the free radiative decay.

On the other hand, 
analyzing photon coincidences under continuous driving for superradiance 
as well as for cooperative emission due to selective measurement, 
we find that both have similar signatures on $g^{(2)}(\infty,\tau)$. 
In particular, in both cases, $g^{(2)}(\infty,\tau)$ has the form of an 
anti-dip, where zero-delay coincidences significantly exceed the limit of two 
independently emitting emitters $g^{(2)}(\infty,0)\le\frac 12$. 
Thus, a violation of this limit is not a unique fingerprint for superradiance.
Instead, it indicates more generally the involvement of correlations between 
emitters during the emission process, which provides the basis for the 
cooperation of both emitters in both photon emission processes.

Furthermore, although specifics of the system like driving and dephasing
generally affect superradiance and cooperative emission due to 
selective measurement differently, the trends are subtle, which prohibits
to clearly distinguish between both underlying mechanism of cooperative 
emission from the measured photon coincidences $g^{(2)}(\infty,\tau)$ alone.

The assessment of time-integrated coincidences 
$\bar{g}^{(2)}_{\Delta\tau}$ for emitters driven by short pulses 
reveals that these can serve as a proxy for instantaneous 
coincidences $g^{(2)}(\infty,0)$ for single and independent identical emitters.
For measurement-induced cooperativity, this only holds 
for small delay-time integration windows $\Delta\tau$,
because dephasing can have a large effect on signals time-integrated over
wider windows.
The situation is even more complex for spontaneously decaying superradiant 
emitters, where time-integration over the non-exponential dynamics 
obfuscates the relation to instantaneous coincidences $g^{(2)}(\infty,0)$.

Summarizing, we find that photon coincidence measurements can signal 
the presence of correlations between emitters in the decay process 
as an anti-dip in $g^{(2)}(t,\tau)$ as a function of $\tau$ 
with values of zero-delay coincidences $g^{(2)}(t,0)$ exceeding the
limit for uncorrelated emitters. However, the origin of these correlations
may be superradiance, cooperative emission without superradiance due to 
emission-angle-selective photon detection, or initial correlations 
induced by the driving. 
To clearly distinguish between these situation, additional information is
required. 
A direct observation of a non-exponential behaviour of the overall 
emitted intensity is likely the most promising strategy 
to unambiguously prove the presence of superradiance.

\acknowledgements
This work was supported by UK EPSRC (grant no. EP/T01377X/1) 
and the ERC (grant no. 725920). 
T.S.S. acknowledges funding by the UK government department for 
Business, Energy and Industrial Strategy through the UK 
national quantum technologies programme.
B.D.G. is supported by a Wolfson Merit Award from the Royal Society 
and a Chair in Emerging Technology from the Royal Academy of Engineering.

\appendix
\section{Microscopic derivation of $\mathbf{k}$-dependent emission and
detection\label{app:krate}}

Here, we show how photon observables for a single photon mode 
$\mathbf{k}$ such as $\langle a^\dagger_{\mathbf{k}}a_{\mathbf{k}}\rangle$
and the corresponding intensity $I_\mathbf{k}$
are related to emitter observables in Markovian emission processes induced
by the light-matter interaction in the Hamiltonian in Eq.~\eqref{eq:hamil}. 
This naturally leads to the picture that the interaction with light field
mode $\mathbf{k}$ gives rise to a decay channel.
From the Heisenberg equations of motion, it follows that

\begin{align}
\frac{\partial}{\partial t}\langle a^\dagger_{\mathbf{k}}a_{\mathbf{k}}\rangle
=&
\frac i\hbar \langle [H, a^\dagger_{\mathbf{k}}a_{\mathbf{k}}]\rangle
\nn
=&-i\sqrt{2}g_{\mathbf{k}} \langle \sigma^-_{\mathbf{k}}a^\dagger_{\mathbf{k}}\rangle
+i\sqrt{2}g_{\mathbf{k}}\langle \sigma^+_{\mathbf{k}} a_{\mathbf{k}}\rangle,
\end{align}
where we have used the definition of $\sigma_\mathbf{k}^\pm$ in 
Eq.~\eqref{eq:defsigmak}.
The light-matter correlations are determined by
\begin{align}
\frac{\partial}{\partial t}
\langle \sigma^-_{\mathbf{k}}a^\dagger_{\mathbf{k}}\rangle
=& i ( \omega_{\mathbf{k}} - \omega )
\langle \sigma^-_{\mathbf{k}}a^\dagger_{\mathbf{k}}\rangle
+i\sqrt{2}g_{\mathbf{k}}\langle [ \sigma^+_{\mathbf{k}} a_{\mathbf{k}},
\sigma^-_{\mathbf{k}}a^\dagger_{\mathbf{k}} ] \rangle
\nn\approx& 
i ( \omega_{\mathbf{k}} - \omega )
\langle \sigma^-_{\mathbf{k}}a^\dagger_{\mathbf{k}}\rangle
+i\sqrt{2}g_{\mathbf{k}}\langle  \sigma^+_{\mathbf{k}} \sigma^-_{\mathbf{k}} \rangle,
\end{align}
where we have simplified 
\begin{align}
\langle [ \sigma^+_{\mathbf{k}} a_{\mathbf{k}},
\sigma^-_{\mathbf{k}}a^\dagger_{\mathbf{k}} ] \rangle
=&
\langle\sigma^+_{\mathbf{k}}\sigma^-_{\mathbf{k}} \rangle
+ \langle [ \sigma^+_{\mathbf{k}}\sigma^-_{\mathbf{k}}- 
            \sigma^-_{\mathbf{k}}\sigma^+_{\mathbf{k}} ] 
a^\dagger_{\mathbf{k}} a_{\mathbf{k}} \rangle
\end{align}
by neglecting the second term involving higher-order contributions.
Integrating the equation of motion for the light-matter correlations yields
\begin{align}
\langle \sigma^-_{\mathbf{k}} a^\dagger_{\mathbf{k}} \rangle=&
i\sqrt{2}g_{\mathbf{k}} \int\limits_{-\infty}^{t}d\tau\, e^{i (\omega_{\mathbf{k}}-\omega)(t-\tau)}
\langle \sigma^+_{\mathbf{k}}\sigma^-_{\mathbf{k}} (\tau) \rangle
\nn
\approx& i\sqrt{2}g_{\mathbf{k}}\pi \delta(\omega_{\mathbf{k}}-\omega) 
\langle \sigma^+_{\mathbf{k}}\sigma^-_{\mathbf{k}} \rangle,
\end{align}
where we have used the Markov approximation 
$\langle \sigma^+_{\mathbf{k}}\sigma^-_{\mathbf{k}} (\tau) \rangle \approx
\langle \sigma^+_{\mathbf{k}}\sigma^-_{\mathbf{k}} (t) \rangle$ and 
neglected the frequency renormalization, i.e. the imaginary part of
\begin{align}
\int\limits_{-\infty}^{t}d\tau\, e^{i (\omega-\omega_{\mathbf{k}})(\tau-t)}
=\pi  \delta (\omega - \omega_{\mathbf{k}}) 
-\frac i{\omega-\omega_{\mathbf{k}}}.
\end{align}

Using
$\langle \sigma^-_{\mathbf{k}}a^\dagger_{\mathbf{k}}\rangle
=\langle \sigma^+_{\mathbf{k}}a_{\mathbf{k}}\rangle^*$,
we arrive at the equation of motion for photon observables
\begin{align}
\frac{\partial}{\partial t}\langle a^\dagger_{\mathbf{k}}a_{\mathbf{k}}\rangle
=& \gamma_\mathbf{k} \langle \sigma^+_{\mathbf{k}}\sigma^-_{\mathbf{k}} \rangle
\label{eq:ddtn_k}
\end{align}
with rate
\begin{align}
\gamma_\mathbf{k}=&2\pi\hbar \, 2 g^2_{\mathbf{k}} \delta(\hbar \omega_{\mathbf{k}}-\hbar \omega).
\label{eq:gammak}
\end{align}

Applying the same steps to the situation of a single emitter, where the 
interaction Hamiltonian is  $ \sum_{\mathbf{k}}\hbar g_\mathbf{k}
(\sigma^- a^\dagger_{\mathbf{k}} + \sigma^+ a_{\mathbf{k}})$, one obtains
a similar result but with a rate 
\begin{align}
\gamma_\mathbf{k}^{\textnormal{single}}=
{2\pi} \hbar g^2_{\mathbf{k}} \delta(\hbar \omega_{\mathbf{k}}-\hbar \omega),
\end{align}
which is half of $\gamma_\mathbf{k}$ in Eq.~\eqref{eq:gammak}.

For two spectrally distinguishable emitters, one has to account for 
two different light-matter correlations 
$\langle \sigma_1^- a^\dagger_{\mathbf{k}}\rangle$
and $\langle \sigma_1^- a^\dagger_{\mathbf{k}}\rangle$ oscillating
with different frequencies $(\omega_{\mathbf{k}}-\omega_1)$ and
$(\omega_{\mathbf{k}}-\omega_2)$, resulting in 
\begin{align}
\frac{\partial}{\partial t}\langle a^\dagger_{\mathbf{k}}a_{\mathbf{k}}\rangle
=& \gamma_\mathbf{k}^{(1)} \langle \sigma^+_1\sigma^-_1 \rangle
+ \gamma_\mathbf{k}^{(2)} \langle \sigma^+_2\sigma^-_2 \rangle,
\label{eq:ddtn_single}
\end{align}
with 
\begin{align}
\gamma_\mathbf{k}^{(i)}=
{2\pi}{\hbar}g^2_{\mathbf{k}} \delta(\hbar \omega_{\mathbf{k}}-\hbar \omega_i),
\end{align}

In addition to the time evolution induced by the light-matter interaction
and described by Eqs.~\eqref{eq:ddtn_k} and Eqs~\eqref{eq:ddtn_single}, 
respectively, the dynamics of the photon mode occupations 
$\langle a^\dagger_\mathbf{k} a_\mathbf{k}\rangle$
is also affected by the destructive measurement at the detectors.
We model the continuous destructive measurement of photons, e.g., using
a single-photon detector\cite{SNSPD}, by a periodic projective measurement of 
$\langle a^\dagger_\mathbf{k} a_\mathbf{k}\rangle$ at time intervals with
width $\Delta \tau_M$ corresponding to the time scale of the measurement, 
which is determined by characteristics of the detector, 
such as the timing jitter. 
The destructive character of the measurement 
implies that the photon state is reset to the vacuum state with
$\langle a^\dagger_\mathbf{k}a_\mathbf{k}\rangle=0$ 
immediately after the measurement, irrespective of the outcome.
The probability of detecting a photon per time interval $\Delta\tau_M$
with a point-like detector measuring photon mode $\mathbf{k}$ is, hence, 
determined by the accumulated excitation transfer from the emitter system
\begin{align}
I_\mathbf{k}(t)=&\frac{1}{\Delta \tau_M} \int\limits_t^{t+\Delta\tau_M} dt'\; 
\bigg(\frac{\partial}{\partial t'} 
\langle a^\dagger_{\mathbf{k}}(t')a_{\mathbf{k}}(t')\rangle \bigg)
\nonumber\\ \approx& 
\frac{\partial}{\partial t}  
\langle a^\dagger_{\mathbf{k}}(t)a_{\mathbf{k}}(t)\rangle
\end{align}
where we have assumed that the source terms, i.e., the emitter occupations
do not change noticably on the time scale $\Delta \tau_M$.
Concretely, for single, two spectrally distinguishable, 
and two indistinguishable emitters, we arrive at
\begin{align}
I_\mathbf{k}^\textnormal{single}(t)=& 
\gamma_\mathbf{k}^{\textnormal{single}} \langle \sigma^+ \sigma^-(t)\rangle,
\\
I_\mathbf{k}^\textnormal{dist}(t)=& 
\gamma_\mathbf{k}^{(1)} \langle \sigma^+_1\sigma^-_1 (t)\rangle
+ \gamma_\mathbf{k}^{(2)} \langle \sigma^+_2\sigma^-_2 (t)\rangle,
\\
I_\mathbf{k}^\textnormal{indist}(t)=&\gamma_{\mathbf{k}} 
\langle \sigma^+_{\mathbf{k}}\sigma^-_{\mathbf{k}}(t)\rangle,
\end{align}
respectively.

\section{Involvement of correlations when $g^{(2)}(t,0)>\frac 12$
\label{app:involveCor}}
We now show why, for a system of two emitters subject to Markovian emission
processes, 
a violation of $g^{(2)}(t,0)\le\frac 12$ indicates the presence of
inter-emitter correlations. 

Note that the derivation of this limit in Eq.~\eqref{eq:glimit} 
is based on two assumptions: i) Absence of initial correlations and 
ii) photon coincidences are given by Eq.~\eqref{eq:G2dist}, which is the
expression for distinguishable emitter.

The former case i) trivially involves correlations. An example 
is a situation with $n_{e_1,g_2}=n_{g_1,e_2}=0$, where 
$g^{(2)}_{\textnormal{dist.}}(t,0)=1/(2{n_{e_1,e_2}(t)})$.
Then, photon coincidences can reach arbitrarily high values when 
$n_{e_1,e_2}$ is small.

As discussed in Appendix~\ref{app:krate}, ii) is valid for Markovian 
emission processes if the emitters are spectrally distinguishable. Thus, for
ii) to break down, the emitters must have the same transition frequencies
$\omega_1=\omega_2$.
Because of the Markovian outcoupling as well as the conservation of the number 
of excitations, photon coincidences must be described as
\begin{align}
G^{(2)}(t,\tau)\propto \sum_{j,j'} \langle \varsigma_j^+(t) 
\varsigma_{j'}^+(t+\tau)\varsigma_{j'}^-(t+\tau)\varsigma_j^-(t)\rangle,
\end{align}
in terms of two pairs of raising and lowering operators $\varsigma_j^\pm$ 
with respect to the emitter state, which are linear combinations 
$\varsigma_j^-=\alpha_j \sigma_1^- + \beta_j \sigma_2^-$ and
$\varsigma_j^+=\alpha^*_j \sigma_1^+ + \beta^*_j \sigma_2^+$ 
of operators $\sigma^\pm_1$ and $\sigma^\pm_2$ for the first and
second emitter, respectively (Compare with Appendix~\ref{app:krate}).
If, for all contributions $j$, either $\alpha_j$ or $\beta_j$ are zero, 
one recovers Eq.~\eqref{eq:G2dist}. Thus, the only situation that remains to
be discussed is that where there is at least one $j$ for which
$\varsigma_j^-$ is a genuine linear combination of $\sigma^-_1$ and 
$\sigma^-_2$ with both $\alpha_j$ and $\beta_j$ nonzero.

Noting also that a non-zero contribution to zero-delay coincidences
$G^{(2)}(t,0)$ are due to occupations of the doubly excited state $n_{e_1,e_2}$
due to conservation of the number of excitations, we observe that 
$\varsigma_j^-$ applied to $|e_1,e_2\rangle$ yields
\begin{align}
\varsigma_j^-|e_1,e_2\rangle =&\alpha_j |g_1,e_2\rangle + \beta_j | e_1, g_2\rangle.
\end{align}
As both $\alpha_j$ and $\beta_j$ are nonzero, the intermediate state
after the emission of a first photon possesses finite inter-emitter correlations
of magnitude $|\alpha_j \beta^*_j|$.

Thus, if $g^{(2)}(t,0)>\frac 12$,
correlations are involved in the emission process in the initial state, 
in the intermediate state after the emission of the first photon, or both.

\section{Continuous incoherent pumping of single and distinguishable emitters
\label{app:single_dist}}
The excited state population $n_i$ of
single emitter subject to radiative decay with rate $\gamma$
and incoherently pumped with a rate $\gamma_p$ can be described 
\begin{align}
\frac{\partial}{\partial t} n_i =&\gamma_p( 1-n_i) -\gamma n_i,
\end{align}
where $( 1-n_i)$ is the population of the ground state.
This equation of motion is solved by
\begin{align}
n_i(t)=&\frac{\gamma_p}{\gamma+\gamma_p}+
\bigg[n_i(0)-\frac{\gamma_p}{\gamma+\gamma_p}\bigg]
e^{-(\gamma+\gamma_p)t}.
\end{align}

Photon coincidences from a single emitter require the detection of a first
photon at time $t$, which occurs with a probability $I_0 n_1(t)$
and implies a collapse of the emitter state onto the ground state, which
defines the initial value $n_1(\tau=0)=0$ 
for the propagation for the delay time $\tau$. Therefore,
\begin{align}
G^{(2)}_\textnormal{single}(t,\tau)=& I_0^2 
\langle \sigma_1^+(t)\sigma_1^+(t+\tau)\sigma_1^-(t+\tau)\sigma_1^-(t)\rangle
\nn=&
I_0^2 n_1(t) \frac{\gamma_p}{\gamma+\gamma_p}
\Big( 1 -e^{-(\gamma+\gamma_p)\tau}\Big).
\end{align}

Evaluating the coincidences from the stationary state $t\to\infty$ and 
normalizing by the squared intensity 
$I_0^2(\infty)=I_0^2 \big[\gamma_p/(\gamma_p+\gamma)\big]^2$,
one obtains
\begin{align}
g^{(2)}_\textnormal{single}(\infty,\tau)=&
1 -e^{-(\gamma+\gamma_p)\tau}.
\end{align}

For two distinguishable emitters, the photon coincidences are
\begin{align}
G^{(2)}_\textnormal{dist}(t,\tau)=&I_0^2 \sum_{i,j=1,2}
\langle \sigma_i^+(t)\sigma_j^+(t+\tau)\sigma_j^-(t+\tau)\sigma_i^-(t)\rangle
\nn=& 
I_0^2 \sum_{i}\langle \sigma_i^+(t)\sigma_i^+(t+\tau)\sigma_i^-(t+\tau)\sigma_i^-(t)\rangle
\nn&+ I_0^2 \sum_{i, j\neq i} \langle \sigma_i^+(t)\sigma_i^-(t)\rangle
\langle \sigma_j^+(t+\tau)\sigma_j^-(t+\tau)\rangle
\nn=& 2 G^{(2)}_\textnormal{single}(t,\tau) + 
2 I_\textnormal{single}(t)  I_\textnormal{single}(t+\tau).
\end{align}
With $I_\textnormal{dist}(t)=2I_\textnormal{single}(t)$, the normalized
coincidences from the stationary state are
\begin{align}
g^{(2)}_\textnormal{dist}(\infty,\tau)=&1 - \frac 12 e^{-(\gamma+\gamma_p)\tau}.
\end{align}

\section{Time-integrated coincidences for superradiant emitters
\label{app:pulsed_sup}}
We consider the free radiative decay of a superradiant two-emitter system 
after optical excitation without additional dephasing.
The equations of motion for the occupations 
of the doubly excited state $n_{e_1,e_2}$ and the symmetric Dicke state 
$n_S$ are
\begin{align}
\frac{\partial}{\partial t} n_{e_1,e_2}=& -\gamma_S n_{e_1,e_2},\\
\frac{\partial}{\partial t} n_S=& \gamma_S (n_{e_1,e_2} -n_S),
\end{align}
with superradiant rate $\gamma_S=2\gamma$.
These equation are solved by
\begin{align}
n_{e_1,e_2}(t)=&n_{e_1,e_2}(0)e^{-\gamma_S t},
\\
n_S(t)=& \big[n_S(0) + \gamma_S t \, n_{e_1,e_2}(0)\big]e^{-\gamma_S t}.
\end{align}
The instantaneous emitted intensity according to Eq.~\eqref{eq:Isup} is 
\begin{align}
I_\textnormal{sup}(t)=& 2I_0 \big[n_{e_1,e_2}(t)+ n_S(t)\big]
\nonumber\\=&
2I_0 \big[n_{e_1,e_2}(0)+n_S(0) + \gamma_S t\, n_{e_1,e_2}(0)\big]
e^{-\gamma_S t}.
\end{align}
With Eqs.~\eqref{eq:G2_10} and \eqref{eq:G2_1T}, we find for the superradiant
photon coincidences defined by Eq.~\eqref{eq:G2sup}
\begin{align}
\bar{G}^{(2)}_{1,\Delta \tau \to 0}=&\Delta \tau 4I^2_0
\frac{\frac 52 n_{e_1,e_2}^2(0)+3n_{e_1,e_2}(0)n_S(0)+ n_S^2(0)}{2\gamma_S}, 
\\
\bar{G}^{(2)}_{1,\Delta \tau \to T}=&4I^2_0 \bigg[
\frac{ 2n_{e_1,e_2}(0) +n_S(0)}{\gamma_S}\bigg]^2.
\end{align}

The only term contributing to coincidences is the initially doubly excited 
state occupation $n_{e_1,e_2}$, which radiatively decays for time $t$, is
then translated into occupations of the symmetric Dicke state $n_S$ by
applications of operators $\sigma^\pm_S$, and subsequent decays from there.
This results in
\begin{align}
\bar{G}^{(2)}_{0,\Delta \tau \to 0}=&\Delta \tau 4I^2_0
\int\limits_0^{T/2}dt\, n_{e_1,e_2}(0) e^{-\gamma_S t} 
\nonumber\\=&
\frac{\Delta \tau 4I^2_0}{\gamma_S} n_{e_1,e_2}(0), 
\\
\bar{G}^{(2)}_{0,\Delta \tau \to T}=& 8 I^2_0
\int\limits_0^{T/2}dt\int\limits_0^{T/2}d\tau\, n_{e_1,e_2}(0) 
e^{-\gamma_S (t+\tau)}
\nonumber\\=&
\frac{8I^2_0}{\gamma_S^2} n_{e_1,e_2}(0).
\end{align}

\section{Time-integrated coincidences for selectively measured emitters
\label{app:pulsed_sel}}
In the case of cooperative emission due to selective measurement with
two identical emitter, the relevant quantities are the occupation
of the doubly excited state $n_{e_1,e_2}$, the occupation of exactly one site
$n_{e_1,g_2}=n_{g_1,e_2}$ as well as the correlations $c$ between 
the states with exactly one excitation.

The equations of motion are
\begin{align}
\frac{\partial}{\partial t}c=&-(\gamma+\gamma_d) c,\\
\frac{\partial}{\partial t}n_{e_1,e_2}=&-2\gamma n_{e_1,e_2},\\
\frac{\partial}{\partial t}n_{e_1,g_2}=&-\gamma n_{e_1,g_2} +\gamma n_{e_1,e_2},
\end{align}
which are solved by
\begin{align}
c(t)=&c(0)e^{-(\gamma+\gamma_d)t}, \\
n_{e_1,e_2}(t)=&n_{e_1,e_2}(0)e^{-2\gamma t}, \\
n_{e_1,g_2}(t)=&
\big[n_{e_1,g_2}(0)+n_{e_1,e_2}(0)\big]e^{-\gamma t}
- n_{e_1,e_2}(0) e^{-2\gamma t}.
\end{align}
The emitted intensity is
\begin{align}
I_\textnormal{sel.}(t)=& 2I_0 \big[ n_{e_1,e_2}(t) + n_{e_1,g_2}(t) +c(t)\big]
\nonumber\\=&
2I_0 \bigg[\big[n_{e_1,g_2}(0)+n_{e_1,e_2}(0)\big]e^{-\gamma t}
+c(0)e^{-(\gamma+\gamma_d)t} \bigg],
\end{align}
which yields
\begin{align}
\bar{G}^{(2)}_{1,\Delta \tau \to 0}=&\Delta \tau 4I^2_0\bigg[
\frac{\big[n_{e_1,g_2}(0)+n_{e_1,e_2}(0)\big]^2}{2\gamma}
\nonumber\\&
+\frac{2\big[n_{e_1,g_2}(0)+n_{e_1,e_2}(0)\big]c(0)}{2\gamma+\gamma_d}
+\frac{c^2(0)}{2\gamma+2\gamma_d}\bigg],
\\
\bar{G}^{(2)}_{1,\Delta \tau \to T}=&4I_0^2
\bigg[\frac{n_{e_1,g_2}(0)+n_{e_1,e_2}(0)}{\gamma}
+\frac{c(0)}{\gamma+\gamma_d}\bigg]^2.
\end{align}

The coincidences are determined by the decay of $n_{e_1,e_2}$ until time $t$,
which is collapsed onto the symmetric Dicke state for which 
$n'_{e_1,g_2}(\tau=0)=c'(\tau=0)=\frac 12n_{e_1,e_2}(t)$. 
These occupations and correlations then decay for time $\tau$. This yields
\begin{align}
\bar{G}^{(2)}_{0,\Delta \tau \to 0}=&
\Delta \tau 4I_0^2 \frac{1}{2\gamma} n_{e_1,e_2}(0),
\\
\bar{G}^{(2)}_{0,\Delta \tau \to T}
=&4I_0^2 \frac{1}{2\gamma}n_{e_1,e_2}(0) \bigg[
\frac{1}{\gamma} + \frac{1}{\gamma+\gamma_d}\bigg].
\end{align}


\begin{thebibliography}{39}%
\makeatletter
\providecommand \@ifxundefined [1]{%
 \@ifx{#1\undefined}
}%
\providecommand \@ifnum [1]{%
 \ifnum #1\expandafter \@firstoftwo
 \else \expandafter \@secondoftwo
 \fi
}%
\providecommand \@ifx [1]{%
 \ifx #1\expandafter \@firstoftwo
 \else \expandafter \@secondoftwo
 \fi
}%
\providecommand \natexlab [1]{#1}%
\providecommand \enquote  [1]{``#1''}%
\providecommand \bibnamefont  [1]{#1}%
\providecommand \bibfnamefont [1]{#1}%
\providecommand \citenamefont [1]{#1}%
\providecommand \href@noop [0]{\@secondoftwo}%
\providecommand \href [0]{\begingroup \@sanitize@url \@href}%
\providecommand \@href[1]{\@@startlink{#1}\@@href}%
\providecommand \@@href[1]{\endgroup#1\@@endlink}%
\providecommand \@sanitize@url [0]{\catcode `\\12\catcode `\$12\catcode
  `\&12\catcode `\#12\catcode `\^12\catcode `\_12\catcode `\%12\relax}%
\providecommand \@@startlink[1]{}%
\providecommand \@@endlink[0]{}%
\providecommand \url  [0]{\begingroup\@sanitize@url \@url }%
\providecommand \@url [1]{\endgroup\@href {#1}{\urlprefix }}%
\providecommand \urlprefix  [0]{URL }%
\providecommand \Eprint [0]{\href }%
\providecommand \doibase [0]{http://dx.doi.org/}%
\providecommand \selectlanguage [0]{\@gobble}%
\providecommand \bibinfo  [0]{\@secondoftwo}%
\providecommand \bibfield  [0]{\@secondoftwo}%
\providecommand \translation [1]{[#1]}%
\providecommand \BibitemOpen [0]{}%
\providecommand \bibitemStop [0]{}%
\providecommand \bibitemNoStop [0]{.\EOS\space}%
\providecommand \EOS [0]{\spacefactor3000\relax}%
\providecommand \BibitemShut  [1]{\csname bibitem#1\endcsname}%
\let\auto@bib@innerbib\@empty
\bibitem [{\citenamefont {{Einstein}}(1916)}]{Einstein}%
  \BibitemOpen
  \bibfield  {author} {\bibinfo {author} {\bibfnamefont {A.}~\bibnamefont
  {{Einstein}}},\ }\href@noop {} {\bibfield  {journal} {\bibinfo  {journal}
  {Verh. Deutsch. Phys. Ges.}\ }\textbf {\bibinfo {volume} {18}},\ \bibinfo
  {pages} {318} (\bibinfo {year} {1916})}\BibitemShut {NoStop}%
\bibitem [{\citenamefont {Weisskopf}(1935)}]{Weisskopf}%
  \BibitemOpen
  \bibfield  {author} {\bibinfo {author} {\bibfnamefont {V.}~\bibnamefont
  {Weisskopf}},\ }\href {\doibase 10.1007/BF01492012} {\bibfield  {journal}
  {\bibinfo  {journal} {Naturwissenschaften}\ }\textbf {\bibinfo {volume}
  {23}},\ \bibinfo {pages} {631} (\bibinfo {year} {1935})}\BibitemShut
  {NoStop}%
\bibitem [{\citenamefont {Drexhage}\ \emph {et~al.}(1968)\citenamefont
  {Drexhage}, \citenamefont {Kuhn},\ and\ \citenamefont
  {Sch{\"a}fer}}]{Drexhage}%
  \BibitemOpen
  \bibfield  {author} {\bibinfo {author} {\bibfnamefont {K.~H.}\ \bibnamefont
  {Drexhage}}, \bibinfo {author} {\bibfnamefont {H.}~\bibnamefont {Kuhn}}, \
  and\ \bibinfo {author} {\bibfnamefont {F.~P.}\ \bibnamefont {Sch{\"a}fer}},\
  }\href {\doibase https://doi.org/10.1002/bbpc.19680720261} {\bibfield
  {journal} {\bibinfo  {journal} {Ber. Bunsenges. Phys. Chem}\ }\textbf
  {\bibinfo {volume} {72}},\ \bibinfo {pages} {329} (\bibinfo {year}
  {1968})}\BibitemShut {NoStop}%
\bibitem [{\citenamefont {Kim}\ \emph {et~al.}(2018)\citenamefont {Kim},
  \citenamefont {Aghaeimeibodi}, \citenamefont {Richardson}, \citenamefont
  {Leavitt},\ and\ \citenamefont {Waks}}]{Waks2018}%
  \BibitemOpen
  \bibfield  {author} {\bibinfo {author} {\bibfnamefont {J.-H.}\ \bibnamefont
  {Kim}}, \bibinfo {author} {\bibfnamefont {S.}~\bibnamefont {Aghaeimeibodi}},
  \bibinfo {author} {\bibfnamefont {C.~J.~K.}\ \bibnamefont {Richardson}},
  \bibinfo {author} {\bibfnamefont {R.~P.}\ \bibnamefont {Leavitt}}, \ and\
  \bibinfo {author} {\bibfnamefont {E.}~\bibnamefont {Waks}},\ }\href {\doibase
  10.1021/acs.nanolett.8b01133} {\bibfield  {journal} {\bibinfo  {journal}
  {Nano Letters}\ }\textbf {\bibinfo {volume} {18}},\ \bibinfo {pages} {4734}
  (\bibinfo {year} {2018})}\BibitemShut {NoStop}%
\bibitem [{\citenamefont {J{\"o}ns}\ \emph {et~al.}(2017)\citenamefont
  {J{\"o}ns}, \citenamefont {Schweickert}, \citenamefont {Versteegh},
  \citenamefont {Dalacu}, \citenamefont {Poole}, \citenamefont {Gulinatti},
  \citenamefont {Giudice}, \citenamefont {Zwiller},\ and\ \citenamefont
  {Reimer}}]{Joens2017}%
  \BibitemOpen
  \bibfield  {author} {\bibinfo {author} {\bibfnamefont {K.~D.}\ \bibnamefont
  {J{\"o}ns}}, \bibinfo {author} {\bibfnamefont {L.}~\bibnamefont
  {Schweickert}}, \bibinfo {author} {\bibfnamefont {M.~A.~M.}\ \bibnamefont
  {Versteegh}}, \bibinfo {author} {\bibfnamefont {D.}~\bibnamefont {Dalacu}},
  \bibinfo {author} {\bibfnamefont {P.~J.}\ \bibnamefont {Poole}}, \bibinfo
  {author} {\bibfnamefont {A.}~\bibnamefont {Gulinatti}}, \bibinfo {author}
  {\bibfnamefont {A.}~\bibnamefont {Giudice}}, \bibinfo {author} {\bibfnamefont
  {V.}~\bibnamefont {Zwiller}}, \ and\ \bibinfo {author} {\bibfnamefont
  {M.~E.}\ \bibnamefont {Reimer}},\ }\href {\doibase
  10.1038/s41598-017-01509-6} {\bibfield  {journal} {\bibinfo  {journal}
  {Scientific Reports}\ }\textbf {\bibinfo {volume} {7}},\ \bibinfo {pages}
  {1700} (\bibinfo {year} {2017})}\BibitemShut {NoStop}%
\bibitem [{\citenamefont {Thomas}\ \emph {et~al.}(2021)\citenamefont {Thomas},
  \citenamefont {Billard}, \citenamefont {Coste}, \citenamefont {Wein},
  \citenamefont {Priya}, \citenamefont {Ollivier}, \citenamefont {Krebs},
  \citenamefont {Taza\"{\i}rt}, \citenamefont {Harouri}, \citenamefont
  {Lemaitre}, \citenamefont {Sagnes}, \citenamefont {Anton}, \citenamefont
  {Lanco}, \citenamefont {Somaschi}, \citenamefont {Loredo},\ and\
  \citenamefont {Senellart}}]{Thomas2021}%
  \BibitemOpen
  \bibfield  {author} {\bibinfo {author} {\bibfnamefont {S.~E.}\ \bibnamefont
  {Thomas}}, \bibinfo {author} {\bibfnamefont {M.}~\bibnamefont {Billard}},
  \bibinfo {author} {\bibfnamefont {N.}~\bibnamefont {Coste}}, \bibinfo
  {author} {\bibfnamefont {S.~C.}\ \bibnamefont {Wein}}, \bibinfo {author}
  {\bibnamefont {Priya}}, \bibinfo {author} {\bibfnamefont {H.}~\bibnamefont
  {Ollivier}}, \bibinfo {author} {\bibfnamefont {O.}~\bibnamefont {Krebs}},
  \bibinfo {author} {\bibfnamefont {L.}~\bibnamefont {Taza\"{\i}rt}}, \bibinfo
  {author} {\bibfnamefont {A.}~\bibnamefont {Harouri}}, \bibinfo {author}
  {\bibfnamefont {A.}~\bibnamefont {Lemaitre}}, \bibinfo {author}
  {\bibfnamefont {I.}~\bibnamefont {Sagnes}}, \bibinfo {author} {\bibfnamefont
  {C.}~\bibnamefont {Anton}}, \bibinfo {author} {\bibfnamefont
  {L.}~\bibnamefont {Lanco}}, \bibinfo {author} {\bibfnamefont
  {N.}~\bibnamefont {Somaschi}}, \bibinfo {author} {\bibfnamefont {J.~C.}\
  \bibnamefont {Loredo}}, \ and\ \bibinfo {author} {\bibfnamefont
  {P.}~\bibnamefont {Senellart}},\ }\href {\doibase
  10.1103/PhysRevLett.126.233601} {\bibfield  {journal} {\bibinfo  {journal}
  {Phys. Rev. Lett.}\ }\textbf {\bibinfo {volume} {126}},\ \bibinfo {pages}
  {233601} (\bibinfo {year} {2021})}\BibitemShut {NoStop}%
\bibitem [{\citenamefont {Cosacchi}\ \emph {et~al.}(2019)\citenamefont
  {Cosacchi}, \citenamefont {Ungar}, \citenamefont {Cygorek}, \citenamefont
  {Vagov},\ and\ \citenamefont {Axt}}]{PI_singlephoton}%
  \BibitemOpen
  \bibfield  {author} {\bibinfo {author} {\bibfnamefont {M.}~\bibnamefont
  {Cosacchi}}, \bibinfo {author} {\bibfnamefont {F.}~\bibnamefont {Ungar}},
  \bibinfo {author} {\bibfnamefont {M.}~\bibnamefont {Cygorek}}, \bibinfo
  {author} {\bibfnamefont {A.}~\bibnamefont {Vagov}}, \ and\ \bibinfo {author}
  {\bibfnamefont {V.~M.}\ \bibnamefont {Axt}},\ }\href {\doibase
  10.1103/PhysRevLett.123.017403} {\bibfield  {journal} {\bibinfo  {journal}
  {Phys. Rev. Lett.}\ }\textbf {\bibinfo {volume} {123}},\ \bibinfo {pages}
  {017403} (\bibinfo {year} {2019})}\BibitemShut {NoStop}%
\bibitem [{\citenamefont {Liu}\ \emph {et~al.}(2018)\citenamefont {Liu},
  \citenamefont {Brash}, \citenamefont {O'Hara}, \citenamefont {Martins},
  \citenamefont {Phillips}, \citenamefont {Coles}, \citenamefont {Royall},
  \citenamefont {Clarke}, \citenamefont {Bentham}, \citenamefont {Prtljaga},
  \citenamefont {Itskevich}, \citenamefont {Wilson}, \citenamefont {Skolnick},\
  and\ \citenamefont {Fox}}]{Fox_Purcell}%
  \BibitemOpen
  \bibfield  {author} {\bibinfo {author} {\bibfnamefont {F.}~\bibnamefont
  {Liu}}, \bibinfo {author} {\bibfnamefont {A.~J.}\ \bibnamefont {Brash}},
  \bibinfo {author} {\bibfnamefont {J.}~\bibnamefont {O'Hara}}, \bibinfo
  {author} {\bibfnamefont {L.~M. P.~P.}\ \bibnamefont {Martins}}, \bibinfo
  {author} {\bibfnamefont {C.~L.}\ \bibnamefont {Phillips}}, \bibinfo {author}
  {\bibfnamefont {R.~J.}\ \bibnamefont {Coles}}, \bibinfo {author}
  {\bibfnamefont {B.}~\bibnamefont {Royall}}, \bibinfo {author} {\bibfnamefont
  {E.}~\bibnamefont {Clarke}}, \bibinfo {author} {\bibfnamefont
  {C.}~\bibnamefont {Bentham}}, \bibinfo {author} {\bibfnamefont
  {N.}~\bibnamefont {Prtljaga}}, \bibinfo {author} {\bibfnamefont {I.~E.}\
  \bibnamefont {Itskevich}}, \bibinfo {author} {\bibfnamefont {L.~R.}\
  \bibnamefont {Wilson}}, \bibinfo {author} {\bibfnamefont {M.~S.}\
  \bibnamefont {Skolnick}}, \ and\ \bibinfo {author} {\bibfnamefont {A.~M.}\
  \bibnamefont {Fox}},\ }\href {\doibase 10.1038/s41565-018-0188-x} {\bibfield
  {journal} {\bibinfo  {journal} {Nature Nanotechnology}\ }\textbf {\bibinfo
  {volume} {13}},\ \bibinfo {pages} {835} (\bibinfo {year} {2018})}\BibitemShut
  {NoStop}%
\bibitem [{\citenamefont {Leistikow}\ \emph {et~al.}(2011)\citenamefont
  {Leistikow}, \citenamefont {Mosk}, \citenamefont {Yeganegi}, \citenamefont
  {Huisman}, \citenamefont {Lagendijk},\ and\ \citenamefont
  {Vos}}]{photonic_crystal3D}%
  \BibitemOpen
  \bibfield  {author} {\bibinfo {author} {\bibfnamefont {M.~D.}\ \bibnamefont
  {Leistikow}}, \bibinfo {author} {\bibfnamefont {A.~P.}\ \bibnamefont {Mosk}},
  \bibinfo {author} {\bibfnamefont {E.}~\bibnamefont {Yeganegi}}, \bibinfo
  {author} {\bibfnamefont {S.~R.}\ \bibnamefont {Huisman}}, \bibinfo {author}
  {\bibfnamefont {A.}~\bibnamefont {Lagendijk}}, \ and\ \bibinfo {author}
  {\bibfnamefont {W.~L.}\ \bibnamefont {Vos}},\ }\href {\doibase
  10.1103/PhysRevLett.107.193903} {\bibfield  {journal} {\bibinfo  {journal}
  {Phys. Rev. Lett.}\ }\textbf {\bibinfo {volume} {107}},\ \bibinfo {pages}
  {193903} (\bibinfo {year} {2011})}\BibitemShut {NoStop}%
\bibitem [{\citenamefont {Iles-Smith}\ \emph {et~al.}(2017)\citenamefont
  {Iles-Smith}, \citenamefont {McCutcheon}, \citenamefont {Nazir},\ and\
  \citenamefont {M{\o}rk}}]{Iles-Smith2017}%
  \BibitemOpen
  \bibfield  {author} {\bibinfo {author} {\bibfnamefont {J.}~\bibnamefont
  {Iles-Smith}}, \bibinfo {author} {\bibfnamefont {D.~P.~S.}\ \bibnamefont
  {McCutcheon}}, \bibinfo {author} {\bibfnamefont {A.}~\bibnamefont {Nazir}}, \
  and\ \bibinfo {author} {\bibfnamefont {J.}~\bibnamefont {M{\o}rk}},\ }\href
  {\doibase 10.1038/nphoton.2017.101} {\bibfield  {journal} {\bibinfo
  {journal} {Nature Photonics}\ }\textbf {\bibinfo {volume} {11}},\ \bibinfo
  {pages} {521} (\bibinfo {year} {2017})}\BibitemShut {NoStop}%
\bibitem [{\citenamefont {del Valle}\ \emph {et~al.}(2012)\citenamefont {del
  Valle}, \citenamefont {Gonzalez-Tudela}, \citenamefont {Laussy},
  \citenamefont {Tejedor},\ and\ \citenamefont {Hartmann}}]{delVallePRL2012}%
  \BibitemOpen
  \bibfield  {author} {\bibinfo {author} {\bibfnamefont {E.}~\bibnamefont {del
  Valle}}, \bibinfo {author} {\bibfnamefont {A.}~\bibnamefont
  {Gonzalez-Tudela}}, \bibinfo {author} {\bibfnamefont {F.~P.}\ \bibnamefont
  {Laussy}}, \bibinfo {author} {\bibfnamefont {C.}~\bibnamefont {Tejedor}}, \
  and\ \bibinfo {author} {\bibfnamefont {M.~J.}\ \bibnamefont {Hartmann}},\
  }\href {\doibase 10.1103/PhysRevLett.109.183601} {\bibfield  {journal}
  {\bibinfo  {journal} {Phys. Rev. Lett.}\ }\textbf {\bibinfo {volume} {109}},\
  \bibinfo {pages} {183601} (\bibinfo {year} {2012})}\BibitemShut {NoStop}%
\bibitem [{\citenamefont {del Valle}\ \emph {et~al.}(2011)\citenamefont {del
  Valle}, \citenamefont {Gonzalez{\textendash}Tudela}, \citenamefont
  {Cancellieri}, \citenamefont {Laussy},\ and\ \citenamefont
  {Tejedor}}]{delValle2011}%
  \BibitemOpen
  \bibfield  {author} {\bibinfo {author} {\bibfnamefont {E.}~\bibnamefont {del
  Valle}}, \bibinfo {author} {\bibfnamefont {A.}~\bibnamefont
  {Gonzalez{\textendash}Tudela}}, \bibinfo {author} {\bibfnamefont
  {E.}~\bibnamefont {Cancellieri}}, \bibinfo {author} {\bibfnamefont {F.~P.}\
  \bibnamefont {Laussy}}, \ and\ \bibinfo {author} {\bibfnamefont
  {C.}~\bibnamefont {Tejedor}},\ }\href {\doibase
  10.1088/1367-2630/13/11/113014} {\bibfield  {journal} {\bibinfo  {journal}
  {New Journal of Physics}\ }\textbf {\bibinfo {volume} {13}},\ \bibinfo
  {pages} {113014} (\bibinfo {year} {2011})}\BibitemShut {NoStop}%
\bibitem [{\citenamefont {Seidelmann}\ \emph {et~al.}(2019)\citenamefont
  {Seidelmann}, \citenamefont {Ungar}, \citenamefont {Cygorek}, \citenamefont
  {Vagov}, \citenamefont {Barth}, \citenamefont {Kuhn},\ and\ \citenamefont
  {Axt}}]{Tim_BiexcitonCascade_PRB}%
  \BibitemOpen
  \bibfield  {author} {\bibinfo {author} {\bibfnamefont {T.}~\bibnamefont
  {Seidelmann}}, \bibinfo {author} {\bibfnamefont {F.}~\bibnamefont {Ungar}},
  \bibinfo {author} {\bibfnamefont {M.}~\bibnamefont {Cygorek}}, \bibinfo
  {author} {\bibfnamefont {A.}~\bibnamefont {Vagov}}, \bibinfo {author}
  {\bibfnamefont {A.~M.}\ \bibnamefont {Barth}}, \bibinfo {author}
  {\bibfnamefont {T.}~\bibnamefont {Kuhn}}, \ and\ \bibinfo {author}
  {\bibfnamefont {V.~M.}\ \bibnamefont {Axt}},\ }\href {\doibase
  10.1103/PhysRevB.99.245301} {\bibfield  {journal} {\bibinfo  {journal} {Phys.
  Rev. B}\ }\textbf {\bibinfo {volume} {99}},\ \bibinfo {pages} {245301}
  (\bibinfo {year} {2019})}\BibitemShut {NoStop}%
\bibitem [{\citenamefont {Seidelmann}\ \emph {et~al.}(2021)\citenamefont
  {Seidelmann}, \citenamefont {Cosacchi}, \citenamefont {Cygorek},
  \citenamefont {Reiter}, \citenamefont {Vagov},\ and\ \citenamefont
  {Axt}}]{Tim_DifferentTypes}%
  \BibitemOpen
  \bibfield  {author} {\bibinfo {author} {\bibfnamefont {T.}~\bibnamefont
  {Seidelmann}}, \bibinfo {author} {\bibfnamefont {M.}~\bibnamefont
  {Cosacchi}}, \bibinfo {author} {\bibfnamefont {M.}~\bibnamefont {Cygorek}},
  \bibinfo {author} {\bibfnamefont {D.~E.}\ \bibnamefont {Reiter}}, \bibinfo
  {author} {\bibfnamefont {A.}~\bibnamefont {Vagov}}, \ and\ \bibinfo {author}
  {\bibfnamefont {V.~M.}\ \bibnamefont {Axt}},\ }\href {\doibase
  https://doi.org/10.1002/qute.202000108} {\bibfield  {journal} {\bibinfo
  {journal} {Advanced Quantum Technologies}\ }\textbf {\bibinfo {volume} {4}},\
  \bibinfo {pages} {2000108} (\bibinfo {year} {2021})}\BibitemShut {NoStop}%
\bibitem [{\citenamefont {Dicke}(1954)}]{Dicke}%
  \BibitemOpen
  \bibfield  {author} {\bibinfo {author} {\bibfnamefont {R.~H.}\ \bibnamefont
  {Dicke}},\ }\href {\doibase 10.1103/PhysRev.93.99} {\bibfield  {journal}
  {\bibinfo  {journal} {Phys. Rev.}\ }\textbf {\bibinfo {volume} {93}},\
  \bibinfo {pages} {99} (\bibinfo {year} {1954})}\BibitemShut {NoStop}%
\bibitem [{\citenamefont {{Gross}}\ and\ \citenamefont
  {{Haroche}}(1982)}]{Haroche}%
  \BibitemOpen
  \bibfield  {author} {\bibinfo {author} {\bibfnamefont {M.}~\bibnamefont
  {{Gross}}}\ and\ \bibinfo {author} {\bibfnamefont {S.}~\bibnamefont
  {{Haroche}}},\ }\href {\doibase 10.1016/0370-1573(82)90102-8} {\bibfield
  {journal} {\bibinfo  {journal} {Phys. Rep}\ }\textbf {\bibinfo {volume}
  {93}},\ \bibinfo {pages} {301} (\bibinfo {year} {1982})}\BibitemShut
  {NoStop}%
\bibitem [{\citenamefont {Bradac}\ \emph {et~al.}(2017)\citenamefont {Bradac},
  \citenamefont {Johnsson}, \citenamefont {Breugel}, \citenamefont {Baragiola},
  \citenamefont {Martin}, \citenamefont {Juan}, \citenamefont {Brennen},\ and\
  \citenamefont {Volz}}]{Bradac2017}%
  \BibitemOpen
  \bibfield  {author} {\bibinfo {author} {\bibfnamefont {C.}~\bibnamefont
  {Bradac}}, \bibinfo {author} {\bibfnamefont {M.~T.}\ \bibnamefont
  {Johnsson}}, \bibinfo {author} {\bibfnamefont {M.~v.}\ \bibnamefont
  {Breugel}}, \bibinfo {author} {\bibfnamefont {B.~Q.}\ \bibnamefont
  {Baragiola}}, \bibinfo {author} {\bibfnamefont {R.}~\bibnamefont {Martin}},
  \bibinfo {author} {\bibfnamefont {M.~L.}\ \bibnamefont {Juan}}, \bibinfo
  {author} {\bibfnamefont {G.~K.}\ \bibnamefont {Brennen}}, \ and\ \bibinfo
  {author} {\bibfnamefont {T.}~\bibnamefont {Volz}},\ }\href {\doibase
  10.1038/s41467-017-01397-4} {\bibfield  {journal} {\bibinfo  {journal}
  {Nature Communications}\ }\textbf {\bibinfo {volume} {8}},\ \bibinfo {pages}
  {1205} (\bibinfo {year} {2017})}\BibitemShut {NoStop}%
\bibitem [{\citenamefont {Higgins}\ \emph {et~al.}(2014)\citenamefont
  {Higgins}, \citenamefont {Benjamin}, \citenamefont {Stace}, \citenamefont
  {Milburn}, \citenamefont {Lovett},\ and\ \citenamefont
  {Gauger}}]{superabsorption_Higgins2014}%
  \BibitemOpen
  \bibfield  {author} {\bibinfo {author} {\bibfnamefont {K.~D.~B.}\
  \bibnamefont {Higgins}}, \bibinfo {author} {\bibfnamefont {S.~C.}\
  \bibnamefont {Benjamin}}, \bibinfo {author} {\bibfnamefont {T.~M.}\
  \bibnamefont {Stace}}, \bibinfo {author} {\bibfnamefont {G.~J.}\ \bibnamefont
  {Milburn}}, \bibinfo {author} {\bibfnamefont {B.~W.}\ \bibnamefont {Lovett}},
  \ and\ \bibinfo {author} {\bibfnamefont {E.~M.}\ \bibnamefont {Gauger}},\
  }\href {\doibase 10.1038/ncomms5705} {\bibfield  {journal} {\bibinfo
  {journal} {Nature Communications}\ }\textbf {\bibinfo {volume} {5}},\
  \bibinfo {pages} {4705} (\bibinfo {year} {2014})}\BibitemShut {NoStop}%
\bibitem [{\citenamefont {Yang}\ \emph {et~al.}(2021)\citenamefont {Yang},
  \citenamefont {Oh}, \citenamefont {Han}, \citenamefont {Son}, \citenamefont
  {Kim}, \citenamefont {Kim}, \citenamefont {Lee},\ and\ \citenamefont
  {An}}]{superabsorption_Yang2021}%
  \BibitemOpen
  \bibfield  {author} {\bibinfo {author} {\bibfnamefont {D.}~\bibnamefont
  {Yang}}, \bibinfo {author} {\bibfnamefont {S.-h.}\ \bibnamefont {Oh}},
  \bibinfo {author} {\bibfnamefont {J.}~\bibnamefont {Han}}, \bibinfo {author}
  {\bibfnamefont {G.}~\bibnamefont {Son}}, \bibinfo {author} {\bibfnamefont
  {J.}~\bibnamefont {Kim}}, \bibinfo {author} {\bibfnamefont {J.}~\bibnamefont
  {Kim}}, \bibinfo {author} {\bibfnamefont {M.}~\bibnamefont {Lee}}, \ and\
  \bibinfo {author} {\bibfnamefont {K.}~\bibnamefont {An}},\ }\href {\doibase
  10.1038/s41566-021-00770-6} {\bibfield  {journal} {\bibinfo  {journal}
  {Nature Photonics}\ }\textbf {\bibinfo {volume} {15}},\ \bibinfo {pages}
  {272} (\bibinfo {year} {2021})}\BibitemShut {NoStop}%
\bibitem [{\citenamefont {Quach}\ \emph {et~al.}(2022)\citenamefont {Quach},
  \citenamefont {McGhee}, \citenamefont {Ganzer}, \citenamefont {Rouse},
  \citenamefont {Lovett}, \citenamefont {Gauger}, \citenamefont {Keeling},
  \citenamefont {Cerullo}, \citenamefont {Lidzey},\ and\ \citenamefont
  {Virgili}}]{quantum_battery}%
  \BibitemOpen
  \bibfield  {author} {\bibinfo {author} {\bibfnamefont {J.~Q.}\ \bibnamefont
  {Quach}}, \bibinfo {author} {\bibfnamefont {K.~E.}\ \bibnamefont {McGhee}},
  \bibinfo {author} {\bibfnamefont {L.}~\bibnamefont {Ganzer}}, \bibinfo
  {author} {\bibfnamefont {D.~M.}\ \bibnamefont {Rouse}}, \bibinfo {author}
  {\bibfnamefont {B.~W.}\ \bibnamefont {Lovett}}, \bibinfo {author}
  {\bibfnamefont {E.~M.}\ \bibnamefont {Gauger}}, \bibinfo {author}
  {\bibfnamefont {J.}~\bibnamefont {Keeling}}, \bibinfo {author} {\bibfnamefont
  {G.}~\bibnamefont {Cerullo}}, \bibinfo {author} {\bibfnamefont {D.~G.}\
  \bibnamefont {Lidzey}}, \ and\ \bibinfo {author} {\bibfnamefont
  {T.}~\bibnamefont {Virgili}},\ }\href {\doibase 10.1126/sciadv.abk3160}
  {\bibfield  {journal} {\bibinfo  {journal} {Science Advances}\ }\textbf
  {\bibinfo {volume} {8}},\ \bibinfo {pages} {eabk3160} (\bibinfo {year}
  {2022})}\BibitemShut {NoStop}%
\bibitem [{\citenamefont {Grim}\ \emph {et~al.}(2019)\citenamefont {Grim},
  \citenamefont {Bracker}, \citenamefont {Zalalutdinov}, \citenamefont
  {Carter}, \citenamefont {Kozen}, \citenamefont {Kim}, \citenamefont {Kim},
  \citenamefont {Mlack}, \citenamefont {Yakes}, \citenamefont {Lee},\ and\
  \citenamefont {Gammon}}]{Gammon2019}%
  \BibitemOpen
  \bibfield  {author} {\bibinfo {author} {\bibfnamefont {J.~Q.}\ \bibnamefont
  {Grim}}, \bibinfo {author} {\bibfnamefont {A.~S.}\ \bibnamefont {Bracker}},
  \bibinfo {author} {\bibfnamefont {M.}~\bibnamefont {Zalalutdinov}}, \bibinfo
  {author} {\bibfnamefont {S.~G.}\ \bibnamefont {Carter}}, \bibinfo {author}
  {\bibfnamefont {A.~C.}\ \bibnamefont {Kozen}}, \bibinfo {author}
  {\bibfnamefont {M.}~\bibnamefont {Kim}}, \bibinfo {author} {\bibfnamefont
  {C.~S.}\ \bibnamefont {Kim}}, \bibinfo {author} {\bibfnamefont {J.~T.}\
  \bibnamefont {Mlack}}, \bibinfo {author} {\bibfnamefont {M.}~\bibnamefont
  {Yakes}}, \bibinfo {author} {\bibfnamefont {B.}~\bibnamefont {Lee}}, \ and\
  \bibinfo {author} {\bibfnamefont {D.}~\bibnamefont {Gammon}},\ }\href
  {\doibase 10.1038/s41563-019-0418-0} {\bibfield  {journal} {\bibinfo
  {journal} {Nature Materials}\ }\textbf {\bibinfo {volume} {18}},\ \bibinfo
  {pages} {963} (\bibinfo {year} {2019})}\BibitemShut {NoStop}%
\bibitem [{\citenamefont {Koong}\ \emph {et~al.}(2022)\citenamefont {Koong},
  \citenamefont {Cygorek}, \citenamefont {Scerri}, \citenamefont {Santana},
  \citenamefont {Park}, \citenamefont {Song}, \citenamefont {Gauger},\ and\
  \citenamefont {Gerardot}}]{SciAdvCoop}%
  \BibitemOpen
  \bibfield  {author} {\bibinfo {author} {\bibfnamefont {Z.~X.}\ \bibnamefont
  {Koong}}, \bibinfo {author} {\bibfnamefont {M.}~\bibnamefont {Cygorek}},
  \bibinfo {author} {\bibfnamefont {E.}~\bibnamefont {Scerri}}, \bibinfo
  {author} {\bibfnamefont {T.~S.}\ \bibnamefont {Santana}}, \bibinfo {author}
  {\bibfnamefont {S.~I.}\ \bibnamefont {Park}}, \bibinfo {author}
  {\bibfnamefont {J.~D.}\ \bibnamefont {Song}}, \bibinfo {author}
  {\bibfnamefont {E.~M.}\ \bibnamefont {Gauger}}, \ and\ \bibinfo {author}
  {\bibfnamefont {B.~D.}\ \bibnamefont {Gerardot}},\ }\href {\doibase
  10.1126/sciadv.abm8171} {\bibfield  {journal} {\bibinfo  {journal} {Science
  Advances}\ }\textbf {\bibinfo {volume} {8}},\ \bibinfo {pages} {eabm8171}
  (\bibinfo {year} {2022})}\BibitemShut {NoStop}%
\bibitem [{Note1()}]{Note1}%
  \BibitemOpen
  \bibinfo {note} {This is because in absence of correlations $\langle \sigma
  ^+_i \sigma ^+_j\sigma ^-_j\sigma ^-_i\rangle =\langle \sigma ^+_i \sigma
  ^-_i\rangle \langle \sigma ^+_j\sigma ^-_j\rangle =n_i n_j$ for $i\protect
  \neq j$ and $\langle \sigma ^+_i \sigma ^+_j\sigma ^-_j\sigma ^-_i\rangle =0$
  for $i=j$. Hence $G^{(2)}(0)=\DOTSB \sum@ \slimits@ _{i,j\protect \neq i} n_i
  n_j=I^2(0)- \DOTSB \sum@ \slimits@ _i n^2_i$ and
  $g^{(2)}(0)=G^{(2)}(0)/I^2(0)= 1-\leavevmode@ifvmode {\setbox \z@ \hbox
  {\mathsurround \z@ $\nulldelimiterspace \z@ \left (\vcenter to\@ne \big@size
  {}\right .$}\box \z@ }\DOTSB \sum@ \slimits@ _i n^2_i\leavevmode@ifvmode
  {\setbox \z@ \hbox {\mathsurround \z@ $\nulldelimiterspace \z@ \left
  )\vcenter to\@ne \big@size {}\right .$}\box \z@ }/\leavevmode@ifvmode
  {\setbox \z@ \hbox {\mathsurround \z@ $\nulldelimiterspace \z@ \left
  (\vcenter to\@ne \big@size {}\right .$}\box \z@ }\DOTSB \sum@ \slimits@ _i
  n_i\leavevmode@ifvmode {\setbox \z@ \hbox {\mathsurround \z@
  $\nulldelimiterspace \z@ \left )\vcenter to\@ne \big@size {}\right .$}\box
  \z@ }^2$. This expression is maximal for equal $n_i$, for which
  $g^{(2)}(0)=1-N/N^2$.}\BibitemShut {Stop}%
\bibitem [{\citenamefont {Wolf}\ \emph {et~al.}(2020)\citenamefont {Wolf},
  \citenamefont {Richter}, \citenamefont {von Zanthier},\ and\ \citenamefont
  {Schmidt-Kaler}}]{Schmidt-KalerPRL}%
  \BibitemOpen
  \bibfield  {author} {\bibinfo {author} {\bibfnamefont {S.}~\bibnamefont
  {Wolf}}, \bibinfo {author} {\bibfnamefont {S.}~\bibnamefont {Richter}},
  \bibinfo {author} {\bibfnamefont {J.}~\bibnamefont {von Zanthier}}, \ and\
  \bibinfo {author} {\bibfnamefont {F.}~\bibnamefont {Schmidt-Kaler}},\ }\href
  {\doibase 10.1103/PhysRevLett.124.063603} {\bibfield  {journal} {\bibinfo
  {journal} {Phys. Rev. Lett.}\ }\textbf {\bibinfo {volume} {124}},\ \bibinfo
  {pages} {063603} (\bibinfo {year} {2020})}\BibitemShut {NoStop}%
\bibitem [{\citenamefont {Richter}\ \emph {et~al.}(2022)\citenamefont
  {Richter}, \citenamefont {Wolf}, \citenamefont {von Zanthier},\ and\
  \citenamefont {Schmidt-Kaler}}]{Schmidt-Kaler2022}%
  \BibitemOpen
  \bibfield  {author} {\bibinfo {author} {\bibfnamefont {S.}~\bibnamefont
  {Richter}}, \bibinfo {author} {\bibfnamefont {S.}~\bibnamefont {Wolf}},
  \bibinfo {author} {\bibfnamefont {J.}~\bibnamefont {von Zanthier}}, \ and\
  \bibinfo {author} {\bibfnamefont {F.}~\bibnamefont {Schmidt-Kaler}},\
  }\href@noop {} {\enquote {\bibinfo {title} {Collective photon emission
  patterns from two atoms in free space},}\ } (\bibinfo {year} {2022}),\
  \Eprint {http://arxiv.org/abs/2202.13678} {arXiv:2202.13678 [quant-ph]}
  \BibitemShut {NoStop}%
\bibitem [{\citenamefont {Skornia}\ \emph {et~al.}(2001)\citenamefont
  {Skornia}, \citenamefont {von Zanthier}, \citenamefont {Agarwal},
  \citenamefont {Werner},\ and\ \citenamefont {Walther}}]{Walther}%
  \BibitemOpen
  \bibfield  {author} {\bibinfo {author} {\bibfnamefont {C.}~\bibnamefont
  {Skornia}}, \bibinfo {author} {\bibfnamefont {J.}~\bibnamefont {von
  Zanthier}}, \bibinfo {author} {\bibfnamefont {G.~S.}\ \bibnamefont
  {Agarwal}}, \bibinfo {author} {\bibfnamefont {E.}~\bibnamefont {Werner}}, \
  and\ \bibinfo {author} {\bibfnamefont {H.}~\bibnamefont {Walther}},\ }\href
  {\doibase 10.1103/PhysRevA.64.063801} {\bibfield  {journal} {\bibinfo
  {journal} {Phys. Rev. A}\ }\textbf {\bibinfo {volume} {64}},\ \bibinfo
  {pages} {063801} (\bibinfo {year} {2001})}\BibitemShut {NoStop}%
\bibitem [{\citenamefont {Bojer}\ and\ \citenamefont {von
  Zanthier}(2022)}]{Zanthier_noninteracting}%
  \BibitemOpen
  \bibfield  {author} {\bibinfo {author} {\bibfnamefont {M.}~\bibnamefont
  {Bojer}}\ and\ \bibinfo {author} {\bibfnamefont {J.}~\bibnamefont {von
  Zanthier}},\ }\href {\doibase 10.1103/PhysRevA.106.053712} {\bibfield
  {journal} {\bibinfo  {journal} {Phys. Rev. A}\ }\textbf {\bibinfo {volume}
  {106}},\ \bibinfo {pages} {053712} (\bibinfo {year} {2022})}\BibitemShut
  {NoStop}%
\bibitem [{\citenamefont {Macovei}\ \emph {et~al.}(2007)\citenamefont
  {Macovei}, \citenamefont {Evers}, \citenamefont {Li},\ and\ \citenamefont
  {Keitel}}]{KeitelPRL}%
  \BibitemOpen
  \bibfield  {author} {\bibinfo {author} {\bibfnamefont {M.}~\bibnamefont
  {Macovei}}, \bibinfo {author} {\bibfnamefont {J.}~\bibnamefont {Evers}},
  \bibinfo {author} {\bibfnamefont {G.-x.}\ \bibnamefont {Li}}, \ and\ \bibinfo
  {author} {\bibfnamefont {C.~H.}\ \bibnamefont {Keitel}},\ }\href {\doibase
  10.1103/PhysRevLett.98.043602} {\bibfield  {journal} {\bibinfo  {journal}
  {Phys. Rev. Lett.}\ }\textbf {\bibinfo {volume} {98}},\ \bibinfo {pages}
  {043602} (\bibinfo {year} {2007})}\BibitemShut {NoStop}%
\bibitem [{\citenamefont {Ficek}\ and\ \citenamefont
  {Swain}(2005)}]{FicekBook}%
  \BibitemOpen
  \bibfield  {author} {\bibinfo {author} {\bibfnamefont {Z.}~\bibnamefont
  {Ficek}}\ and\ \bibinfo {author} {\bibfnamefont {S.}~\bibnamefont {Swain}},\
  }\href {\doibase 10.1007/b100106} {\emph {\bibinfo {title} {Quantum
  Interference and Coherence}}},\ edited by\ \bibinfo {editor} {\bibfnamefont
  {W.~T.}\ \bibnamefont {Rhodes}}, \bibinfo {editor} {\bibfnamefont
  {T.}~\bibnamefont {Asakura}}, \bibinfo {editor} {\bibfnamefont {K.-H.}\
  \bibnamefont {Brenner}}, \bibinfo {editor} {\bibfnamefont {T.~W.}\
  \bibnamefont {H\"{a}nsch}}, \bibinfo {editor} {\bibfnamefont
  {T.}~\bibnamefont {Kamiya}}, \bibinfo {editor} {\bibfnamefont
  {F.}~\bibnamefont {Krausz}}, \bibinfo {editor} {\bibfnamefont
  {B.}~\bibnamefont {Monemar}}, \bibinfo {editor} {\bibfnamefont
  {H.}~\bibnamefont {Venghaus}}, \bibinfo {editor} {\bibfnamefont
  {H.}~\bibnamefont {Weber}}, \ and\ \bibinfo {editor} {\bibfnamefont
  {H.}~\bibnamefont {Weinfurter}}\ (\bibinfo  {publisher} {Springer New York},\
  \bibinfo {year} {2005})\BibitemShut {NoStop}%
\bibitem [{\citenamefont {Lax}(1963)}]{QRT_Lax}%
  \BibitemOpen
  \bibfield  {author} {\bibinfo {author} {\bibfnamefont {M.}~\bibnamefont
  {Lax}},\ }\href {\doibase 10.1103/PhysRev.129.2342} {\bibfield  {journal}
  {\bibinfo  {journal} {Phys. Rev.}\ }\textbf {\bibinfo {volume} {129}},\
  \bibinfo {pages} {2342} (\bibinfo {year} {1963})}\BibitemShut {NoStop}%
\bibitem [{\citenamefont {Cosacchi}\ \emph {et~al.}(2021)\citenamefont
  {Cosacchi}, \citenamefont {Seidelmann}, \citenamefont {Cygorek},
  \citenamefont {Vagov}, \citenamefont {Reiter},\ and\ \citenamefont
  {Axt}}]{QRT}%
  \BibitemOpen
  \bibfield  {author} {\bibinfo {author} {\bibfnamefont {M.}~\bibnamefont
  {Cosacchi}}, \bibinfo {author} {\bibfnamefont {T.}~\bibnamefont
  {Seidelmann}}, \bibinfo {author} {\bibfnamefont {M.}~\bibnamefont {Cygorek}},
  \bibinfo {author} {\bibfnamefont {A.}~\bibnamefont {Vagov}}, \bibinfo
  {author} {\bibfnamefont {D.~E.}\ \bibnamefont {Reiter}}, \ and\ \bibinfo
  {author} {\bibfnamefont {V.~M.}\ \bibnamefont {Axt}},\ }\href {\doibase
  10.1103/PhysRevLett.127.100402} {\bibfield  {journal} {\bibinfo  {journal}
  {Phys. Rev. Lett.}\ }\textbf {\bibinfo {volume} {127}},\ \bibinfo {pages}
  {100402} (\bibinfo {year} {2021})}\BibitemShut {NoStop}%
\bibitem [{\citenamefont {Ficek}\ \emph {et~al.}(1987)\citenamefont {Ficek},
  \citenamefont {Tanaś},\ and\ \citenamefont {Kielich}}]{Ficek}%
  \BibitemOpen
  \bibfield  {author} {\bibinfo {author} {\bibfnamefont {Z.}~\bibnamefont
  {Ficek}}, \bibinfo {author} {\bibfnamefont {R.}~\bibnamefont {Tanaś}}, \
  and\ \bibinfo {author} {\bibfnamefont {S.}~\bibnamefont {Kielich}},\ }\href
  {\doibase https://doi.org/10.1016/0378-4371(87)90280-9} {\bibfield  {journal}
  {\bibinfo  {journal} {Physica A: Statistical Mechanics and its Applications}\
  }\textbf {\bibinfo {volume} {146}},\ \bibinfo {pages} {452} (\bibinfo {year}
  {1987})}\BibitemShut {NoStop}%
\bibitem [{\citenamefont {Hanbury~Brown}\ and\ \citenamefont
  {Twiss}(1954)}]{HBT}%
  \BibitemOpen
  \bibfield  {author} {\bibinfo {author} {\bibfnamefont {R.}~\bibnamefont
  {Hanbury~Brown}}\ and\ \bibinfo {author} {\bibfnamefont {R.~Q.}\ \bibnamefont
  {Twiss}},\ }\href {\doibase 10.1080/14786440708520475} {\bibfield  {journal}
  {\bibinfo  {journal} {Philos. Mag.}\ }\textbf {\bibinfo {volume} {45}},\
  \bibinfo {pages} {663} (\bibinfo {year} {1954})}\BibitemShut {NoStop}%
\bibitem [{\citenamefont {Schofield}\ \emph {et~al.}(2022)\citenamefont
  {Schofield}, \citenamefont {Clear}, \citenamefont {Hoggarth}, \citenamefont
  {Major}, \citenamefont {McCutcheon},\ and\ \citenamefont
  {Clark}}]{Clark2022}%
  \BibitemOpen
  \bibfield  {author} {\bibinfo {author} {\bibfnamefont {R.~C.}\ \bibnamefont
  {Schofield}}, \bibinfo {author} {\bibfnamefont {C.}~\bibnamefont {Clear}},
  \bibinfo {author} {\bibfnamefont {R.~A.}\ \bibnamefont {Hoggarth}}, \bibinfo
  {author} {\bibfnamefont {K.~D.}\ \bibnamefont {Major}}, \bibinfo {author}
  {\bibfnamefont {D.~P.~S.}\ \bibnamefont {McCutcheon}}, \ and\ \bibinfo
  {author} {\bibfnamefont {A.~S.}\ \bibnamefont {Clark}},\ }\href {\doibase
  10.1103/PhysRevResearch.4.013037} {\bibfield  {journal} {\bibinfo  {journal}
  {Phys. Rev. Research}\ }\textbf {\bibinfo {volume} {4}},\ \bibinfo {pages}
  {013037} (\bibinfo {year} {2022})}\BibitemShut {NoStop}%
\bibitem [{\citenamefont {Otten}\ \emph {et~al.}(2020)\citenamefont {Otten},
  \citenamefont {Kenneweg}, \citenamefont {Hensen}, \citenamefont {Gray},\ and\
  \citenamefont {Pfeiffer}}]{Otten2020}%
  \BibitemOpen
  \bibfield  {author} {\bibinfo {author} {\bibfnamefont {M.}~\bibnamefont
  {Otten}}, \bibinfo {author} {\bibfnamefont {T.}~\bibnamefont {Kenneweg}},
  \bibinfo {author} {\bibfnamefont {M.}~\bibnamefont {Hensen}}, \bibinfo
  {author} {\bibfnamefont {S.~K.}\ \bibnamefont {Gray}}, \ and\ \bibinfo
  {author} {\bibfnamefont {W.}~\bibnamefont {Pfeiffer}},\ }\href {\doibase
  10.1103/PhysRevA.102.043118} {\bibfield  {journal} {\bibinfo  {journal}
  {Phys. Rev. A}\ }\textbf {\bibinfo {volume} {102}},\ \bibinfo {pages}
  {043118} (\bibinfo {year} {2020})}\BibitemShut {NoStop}%
\bibitem [{\citenamefont {Kiraz}\ \emph {et~al.}(2004)\citenamefont {Kiraz},
  \citenamefont {Atat\"ure},\ and\ \citenamefont {Imamo\ifmmode~\breve{g}\else
  \u{g}\fi{}lu}}]{Kiraz}%
  \BibitemOpen
  \bibfield  {author} {\bibinfo {author} {\bibfnamefont {A.}~\bibnamefont
  {Kiraz}}, \bibinfo {author} {\bibfnamefont {M.}~\bibnamefont {Atat\"ure}}, \
  and\ \bibinfo {author} {\bibfnamefont {A.}~\bibnamefont
  {Imamo\ifmmode~\breve{g}\else \u{g}\fi{}lu}},\ }\href {\doibase
  10.1103/PhysRevA.69.032305} {\bibfield  {journal} {\bibinfo  {journal} {Phys.
  Rev. A}\ }\textbf {\bibinfo {volume} {69}},\ \bibinfo {pages} {032305}
  (\bibinfo {year} {2004})}\BibitemShut {NoStop}%
\bibitem [{\citenamefont {Stevenson}\ \emph {et~al.}(2008)\citenamefont
  {Stevenson}, \citenamefont {Hudson}, \citenamefont {Bennett}, \citenamefont
  {Young}, \citenamefont {Nicoll}, \citenamefont {Ritchie},\ and\ \citenamefont
  {Shields}}]{Stevenson08}%
  \BibitemOpen
  \bibfield  {author} {\bibinfo {author} {\bibfnamefont {R.~M.}\ \bibnamefont
  {Stevenson}}, \bibinfo {author} {\bibfnamefont {A.~J.}\ \bibnamefont
  {Hudson}}, \bibinfo {author} {\bibfnamefont {A.~J.}\ \bibnamefont {Bennett}},
  \bibinfo {author} {\bibfnamefont {R.~J.}\ \bibnamefont {Young}}, \bibinfo
  {author} {\bibfnamefont {C.~A.}\ \bibnamefont {Nicoll}}, \bibinfo {author}
  {\bibfnamefont {D.~A.}\ \bibnamefont {Ritchie}}, \ and\ \bibinfo {author}
  {\bibfnamefont {A.~J.}\ \bibnamefont {Shields}},\ }\href {\doibase
  10.1103/PhysRevLett.101.170501} {\bibfield  {journal} {\bibinfo  {journal}
  {Phys. Rev. Lett.}\ }\textbf {\bibinfo {volume} {101}},\ \bibinfo {pages}
  {170501} (\bibinfo {year} {2008})}\BibitemShut {NoStop}%
\bibitem [{\citenamefont {Cygorek}\ \emph {et~al.}(2018)\citenamefont
  {Cygorek}, \citenamefont {Ungar}, \citenamefont {Seidelmann}, \citenamefont
  {Barth}, \citenamefont {Vagov}, \citenamefont {Axt},\ and\ \citenamefont
  {Kuhn}}]{ConcurrenceCygorek}%
  \BibitemOpen
  \bibfield  {author} {\bibinfo {author} {\bibfnamefont {M.}~\bibnamefont
  {Cygorek}}, \bibinfo {author} {\bibfnamefont {F.}~\bibnamefont {Ungar}},
  \bibinfo {author} {\bibfnamefont {T.}~\bibnamefont {Seidelmann}}, \bibinfo
  {author} {\bibfnamefont {A.~M.}\ \bibnamefont {Barth}}, \bibinfo {author}
  {\bibfnamefont {A.}~\bibnamefont {Vagov}}, \bibinfo {author} {\bibfnamefont
  {V.~M.}\ \bibnamefont {Axt}}, \ and\ \bibinfo {author} {\bibfnamefont
  {T.}~\bibnamefont {Kuhn}},\ }\href {\doibase 10.1103/PhysRevB.98.045303}
  {\bibfield  {journal} {\bibinfo  {journal} {Phys. Rev. B}\ }\textbf {\bibinfo
  {volume} {98}},\ \bibinfo {pages} {045303} (\bibinfo {year}
  {2018})}\BibitemShut {NoStop}%
\bibitem [{\citenamefont {Natarajan}\ \emph {et~al.}(2012)\citenamefont
  {Natarajan}, \citenamefont {Tanner},\ and\ \citenamefont {Hadfield}}]{SNSPD}%
  \BibitemOpen
  \bibfield  {author} {\bibinfo {author} {\bibfnamefont {C.~M.}\ \bibnamefont
  {Natarajan}}, \bibinfo {author} {\bibfnamefont {M.~G.}\ \bibnamefont
  {Tanner}}, \ and\ \bibinfo {author} {\bibfnamefont {R.~H.}\ \bibnamefont
  {Hadfield}},\ }\href {\doibase 10.1088/0953-2048/25/6/063001} {\bibfield
  {journal} {\bibinfo  {journal} {Superconductor Science and Technology}\
  }\textbf {\bibinfo {volume} {25}},\ \bibinfo {pages} {063001} (\bibinfo
  {year} {2012})}\BibitemShut {NoStop}%
\end{thebibliography}
%
 
\end{document}